\documentclass[
superscriptaddress,
pra,
showpacs,
showkeys,
twocolumn,
aps,
floatfix,
]{revtex4-1}

\usepackage{amsmath}
\usepackage{amssymb}

\usepackage{graphicx}
\usepackage{bm,bbm}
\usepackage{color}
\usepackage{placeins}
\usepackage[normalem]{ulem}
\usepackage{soul}
\usepackage{xcolor}

\newcommand{\eg}{\textit{e.g.}}

\newcommand{\rmw}{\mathrm{w}}
\newcommand{\rmd}{\mathrm{d}}
\newcommand{\rme}{\mathrm{e}}
\newcommand{\rmi}{\mathrm{i}}

\newcommand{\ie}{\textit{i.e.} }
\newcommand{\mt}{\mathrm }

\newcommand{\omo}{\omega_{\mathrm{o}}}
\newcommand{\gammo}{\gamma_{\mathrm{o}}}

\newcommand{\One}{\openone}

\newcommand{\h}{\hat}

\newcommand{\la}{\langle}
\newcommand{\ra}{\rangle}

\newcommand{\bR}{\bm{R}}

\begin{document}
\title{Monitoring the dynamics of an open quantum system via a single qubit}

\author{P. C. L\' opez V\' azquez}
\affiliation{Departamento de Ciencias Naturales y Exactas, Universidad de Guadalajara,  
Carretera Guadalajara - Ameca Km. 45.5 C.P. 46600. Ameca, Jalisco, M\'exico.}

\author{T. Gorin}
\affiliation{Departamento de F\'\i sica, Universidad de Guadalajara,
Blvd. Marcelino Garc\'\i a Barragan y Calzada Ol\'\i mpica,   
C.P. 44840, Guadalajara, Jalisco, M\' exico.}

\begin{abstract}
We investigate the possibility to monitor the dynamics of an open quantum system
with the help of a small probe system, coupled via dephasing coupling to the 
open system of interest.  
As an example, we consider a dissipative harmonic oscillator and a single qubit
as probe system. Qubit plus oscillator are described by a finite temperature 
quantum master equation, where the dynamics of the whole system can be obtained
analytically.
We find that the short time behavior of the reduced qubit state (its 
coherence) provides exhaustive information on the dissipative dynamics of the 
oscillator. Observing this coherence for two initial states with different 
out-of-equilibrium temperatures, one can determine all coupling constants and 
the equilibrium temperature fixed by the external heat bath. 
In addition, the dephasing coupling to the qubit probe, may be considered 
as a perturbation of the dissipative oscillator. The corresponding quantum 
fidelity can be calculated analytically, also. Hence, we find the precise 
relation between the behavior of the reduced qubit state (its coherence) and 
that fidelity.
\end{abstract}

\maketitle

\section{\label{I} Introduction}
The idea of probing the dynamics of a quantum system by another smaller quantum
system coupled to the first one, goes probably back to Gardiner, Cirac and 
Zoller~\cite{GCZ97}. In that paper, the authors propose to study the stability 
of the unitary dynamics of a complex, eventually quantum chaotic system, the 
delta-kicked harmonic oscillator, using a probe degree of freedom coupled to
the system by dephasing. That stability is characterized in terms of the 
quantum fidelity (``quantum Loschmidt 
echo'')~\cite{Peres84,Pasta00,GPSZ06,JacPet09}.

More recently, quantum thermodynamics has drawn a lot of attention in part due 
to the difficulties which occur when one tries to extend
the classical thermodynamic concepts such as work, heat, entropy, and 
thermalization to small microscopic quantum systems~\cite{Bi15,Ng15,Mi16,Go16}.
There, it is of fundamental interest to develop accurate techniques for the
verification of thermodynamical properties. In this sense, ``quantum 
thermometry'' has been formulated for single qubit readouts~\cite{Jev15,Man17}.
Further work in this direction has been centered on the construction of quantum 
heat machines~\cite{Qua07,Ra15,Ho13,An17,Hofer17}.

The purpose of the present work consists in extending the original scheme for
probing quantum fidelity to the case of dissipative dynamics. Thereby, we want
to understand how to extract as much information as possible about the 
dissipative dynamics in question. For the quantum chaotic case, some results 
have been obtained in Ref.~\cite{GMS16RTSA} in the case of an infinite 
temperature bath. Here, we are interested in the case of finite temperature and 
a finite coupling strength (dissipation rate), and instead of a quantum chaotic
system as in Ref.~\cite{PLG17}, we study a simple harmonic oscillator. This 
allows us to obtain the dynamics of the full system analytically, and thereby 
study the relations between the dynamics of the oscillator and that of the 
quantum probe in every detail.

Dephasing coupling has been studied in many different 
contexts~\cite{Palma96,Reina02,vanderWal2003,Brito15,Costa16}. The 
qubit-oscillator system with dephasing coupling could be implemented 
experimentally, with superconducting quantum devices~\cite{Wa04,Ch04,Harris10},
trapped ions~\cite{Fri08,Porras08,Kim11,Sch13}, ultracold atoms in an optical 
lattice~\cite{Sim11,Recati05,Orth08,Haikka11}, Josephson 
junctions~\cite{Ma01,So04}, or defect centers in solid-state 
crystals~\cite{BeToBi14}.

The dissipative harmonic oscillator has the extraordinary feature that Gaussian
wave packets continue to evolve as Gaussian wave packets for all times. 
Following Refs.~\cite{Scu98,Isar09} this allows us to compute 
Uhlmann's~\cite{Uhl76} (Jozsa's~\cite{Joz94}) fidelity for mixed quantum states 
in analytical form. We then compare the generalized fidelity which has been
introduced in Ref.~\cite{GMS16RTSA} and is based on the qubit coherence, with 
the standard fidelity for mixed quantum states. In 
Refs.~\cite{Clerk07,Zhao14,Vin15} discuss different possibilities to extract
information about the dynamics of an oscillator with the help of coupled to the
system.

The paper is organized as follows: In Sec.~\ref{model} the details about our 
tripartite model, together with the definitions of the generalized fidelity and 
the Uhlmann-Jozsa fidelity are given. Also we give a short review about the 
Wigner function description and state some of the properties of its 
two dimensional Fourier transform or the chord function. In Sec.~\ref{ansols},
we derive an analytic solution for the dynamics of our model system, and 
discuss the reduced dynamics of the oscillator and the qubit. In 
Sec.~\ref{fidelities}, we find analytical expressions for the generalized 
fidelity and the Uhlmann-Josza fidelity, as well as the connection formulas between
the two fidelities and the purities of the qubit and the oscillator;
furthermore, we use our previous results to propose a method to 
implement a quantum thermometer by looking only at the decoherence decaying 
rate of the qubit. Finally in Sec.~\ref{con} we give our conclusions.

\section{\label{model} The tripartite system}
The system is composed of three parts; a central two-level system (qubit), 
an intermediate harmonic oscillator and a heat bath of finite temperature, 
whose effect is described by a quantum master equation of Lindblad form. 
Assuming a quantum optical setting, and measuring energy in units of 
$\hbar \omo$, the energy quantum of the oscillator, and time in units of 
$\omo^{-1}$, we may write the master equation in terms of dimensionless 
quantities
\begin{equation}\label{meL}
\rmi\, \frac{\rmd\varrho}{\rmd t} = [H, \varrho]
 + \rmi \mathcal{L}[\varrho] \; ,
\end{equation}
where the density matrix $\varrho$ represents the mixed state of qubit
plus oscillator mode. The Hamiltonian is divided into the qubit part,
the oscillator part $H_{\rm osc}$, and the coupling between both systems:
\begin{align}\label{IntroHam}
H &= {\Delta\over 2}\, \sigma_z  + H_{\rm osc} 
         + g \, \sigma_z \otimes \h{x}\; , \\
H_{\rm osc} &= \frac{1}{2}\, \left(\h{x}^2 + \h{p}^2 \right)
   = \hat a^\dagger \hat a + \frac{\One}{2}\; .
\end{align}
The mixed state of the qubit alone is obtained from $\varrho$ via the partial 
trace. In our model, the coupling is of the dephasing type. Therefore, the 
populations of the qubit states are constant in time, and the non-diagonal 
element of the qubit density matrix (``coherence'') is the quantity of 
interest, as it contains all the information about the dynamics of the 
oscillator. The Lindblad term, which accounts for the dissipative processes, is 
given by
\begin{eqnarray}\nonumber
\mathcal{L}[\varrho] = - \kappa \left(1+\bar{n}\right)\left(a^{\dag}a\, \varrho 
- 2\,a\varrho\,a^{\dag} + \varrho\, a^{\dag} a\right)\\\label{Lindo}
- \kappa\, \bar{n}\left(aa^{\dag}\, \varrho - 2\,a^{\dag}\varrho\,a + \varrho\, a a^{\dag}\right) \; ,
\end{eqnarray}
where $\kappa= \gammo/\omo$. Here, 
\begin{equation}
 \bar n = \la \hat a^\dagger \hat a\ra_{ \varrho_T}
   = \frac{1}{\rme ^{1/D} - 1}\; , 
\label{defnbar}\end{equation}
is the average number of excitations, and $\varrho_T$ the canonical 
equilibrium state of the harmonic oscillator at temperature $T$. The 
parameter $D = {k_{\rm B}\, T / (\hbar\, \omo)}$ is the dimensionless diffusion 
constant from the quantum Brownian motion model~\cite{CalLeg83}.

\subsection{\label{mama} Fidelities}
The density operator $\varrho$, which appears in Eq.~(\ref{meL}), describes the 
mixed quantum state of the bipartite system consisting of two-level system 
(qubit) and harmonic oscillator. It may be written in block-matrix form as 
follows:
\begin{equation}
\varrho(t) = \begin{pmatrix} 
    a_{00}\,  \varrho_{00}(t) & a_{01}\,  \varrho_{01}(t)\\
    a_{10}\,  \varrho_{10}(t) & a_{11}\,  \varrho_{11}(t) \end{pmatrix}\; , 
\label{mama:defrho}\end{equation}
where the coefficients $a_{ij}$ are related to the initial state of the qubit
(see below). Each operator $\varrho_{ij}(t)$ acts on the Hilbert space of
the harmonic oscillator. In this way, Eq.~(\ref{meL}) separates into 
independent evolution equations for each of these operators. With 
$H_\pm = H_{\mt{osc}} \pm g\, \h{x}$, we find
\begin{align}
\rmi\, \frac{\rmd \varrho_{00}}{\rmd t} &=
     [H_+ \, ,\,  \varrho_{00}] + \mathcal{L}[ \varrho_{00}]\; , \label{me00}\\
\rmi\, \frac{\rmd \varrho_{11}}{\rmd t} &=
     [H_- \, ,\,  \varrho_{11}] + \mathcal{L}[ \varrho_{11}]\; , \label{me11}\\
\rmi\, \frac{\rmd \varrho_{01}}{\rmd t} &=
     \big (\, H_+\, \varrho_{01} - \varrho_{01}\, H_-\, \big ) 
   + \frac{\Delta}{2}\; \varrho_{01} + \mathcal{L}[\varrho_{01}] \; .
\label{me01}\end{align}
We assume the initial state to be a product state of the form
\begin{align}
\varrho(0) = \begin{pmatrix} a_{00} & a_{01}\\
                 a_{10} & a_{11}\end{pmatrix} \otimes  \varrho_{\mt{osc}} \; . 
\label{mama:initstate}\end{align}
In the evolution equations~(\ref{me00}-\ref{me01}), the coupling term 
between qubit and oscillator appears as a perturbation to the dynamics of the 
oscillator mode. This makes it possible to study its fidelity or (quantum 
Loschmidt echo)~\cite{GPSZ06}. Without dissipation and for a pure initial 
state, this fidelity $F(t)$ can be obtained from both, the diagonal and the 
off-diagonal blocks~\cite{GCZ97,GPSS04}. From the diagonal blocks, we obtain
\begin{align}
\varrho_{00}(t) &= |\psi_+(t)\ra\la\psi_+(t)|\;\; :\;\; 
\psi_+(t)= \rme^{-\rmi H_+ t}\, |\psi(0)\ra\; , \notag\\
\varrho_{11}(t) &= |\psi_-(t)\ra\la\psi_-(t)|\;\; :\;\; 
\psi_-(t)= \rme^{-\rmi H_- t}\, |\psi(0)\ra\; ,
\end{align}
where $|\psi_+(0)\ra = |\psi_-(0)\ra = |\psi(0)\ra$ is the pure initial state 
of the oscillator mode. From this, we obtain the quantum fidelity as 
\begin{align}
F(t)= {\rm Tr}\big [\, \varrho_{00}(t)\, \varrho_{11}(t)\, \big ]
 = \big |\, \la\psi_+(t)| \psi_-(t)\ra \, \big |^2 \; .
\label{mama:diagFid}\end{align}
From the off-diagonal block, we get
\begin{align}
\varrho_{01}(t) &= \rme^{\rmi \Delta t}\; 
   \rme^{-\rmi H_- t}\; \varrho_{01}(0)\; \rme^{\rmi H_+ t} \notag\\
 &= \rme^{\rmi \Delta t}\; |\psi_-(t)\ra\, \la\psi_+(t)| \; ,
\end{align}
while $\varrho_{10}(t)= \varrho_{01}(t)^\dagger$. This allows us to write
\begin{align}
F(t) = {\rm Tr}\big [\, \varrho_{01}(t)\, \varrho_{10}(t)\, \big ] \; .
\label{mama:offdFid}\end{align} 
If we include dissipation and/or mixed initial states, then
the strict equivalence between the Eqs.~(\ref{mama:diagFid}) 
and~(\ref{mama:offdFid}) breaks down. In that case, the operators
$\varrho_{00}(t)$ and $\varrho_{11}(t)$ become density matrices, which are the
solutions of a quantum master equation of Lindblad 
from~\cite{Sud61,GoKoSu76,Lin76}. 
Concerning the diagonal blocks, we use a standard generalization of fidelity 
to the case of mixed quantum states, which is due to Uhlmann~\cite{Uhl76} 
(mathematical definition) and Jozsa~\cite{Joz94}. Thus we define
\begin{equation}\label{uhlf}
 F_{\rm UJ}(t) = {\rm Tr}\big (\,  \varrho_{00}(t)^{1/4}\,
       \varrho_{11}(t)^{1/2}\,  \varrho_{00}(t)^{1/4}\, \big )^2 \; .
\end{equation}      
Concerning the non-diagonal blocks, we interpret Eq.~(\ref{mama:offdFid}) as a 
different measure for (the loss of) fidelity in a open quantum system, and 
denote that quantity
\begin{equation}\label{genf}
F_{\rm gen}(t) =  
   {\rm Tr}\big [\, \varrho_{01}(t)\, \varrho_{10}(t)\, \big ]
 = \left | {\rm Tr}\left[ 
   \varrho_{01}(t) \right]\,  \right |^2 \; .
\end{equation}
as the generalized fidelity~\cite{MGS15,GMS16RTSA}.

There are important conceptual differences between $F_{\rm UJ}(t)$ and 
$F_{\rm gen}(t)$: $F_{\rm UJ}(t)$ can be used to quantify the similarity of 
mixed quantum states, it is not necessary that these are states evolving under 
certain evolution equations. By contrast, $F_{\rm gen}(t)$ requires 
to specify these evolution equations. It also requires that these are 
of Lindblad form and differ in the Hamiltonian part only. Furthermore, in a
typical case, the master equations for the two diagonal blocks guide any 
initial state to the same equilibrium state. Therefore, $F_{\rm UJ}(t)$ will 
typically increase towards one at the end. By contrast, $F_{\rm gen}(t)$ will 
often drop to zero.

\subsection{\label{chf} Wigner and chord function description}
The solutions derived here are carried out by employing the chord function
description~\cite{Ozo98,Ozo02,Oz04}. The chord function (or the 
characteristic Wigner function) is defined as the Fourier transform of the 
Wigner function~\cite{BrePet02}. In what follows we review some of their 
properties.\\

\paragraph{Wigner function} We start with the position representation of an 
operator $\hat A$. If this operator has a matrix representation with respect to
some orthonormal basis $\{ \varphi_j \}_{j\in\mathbb{N}}$,
\begin{align}
\la x|\hat A|x'\ra = \sum_{ij} A_{ij}\; \la x|\varphi_i\ra\; \la\varphi_j|x'\ra
 = \sum_{ij} A_{ij}\; \varphi_i(x)\; \varphi_j(x')^* \; ,
\label{OpPosiRep}\end{align}
where $\la x|\varphi_j\ra$ is the Dirac notation for the familiar wave function 
representation $\varphi_j(x)$. Then, we may define the Wigner function of a 
given collection of quantum states described by the density matrix $\varrho$ as
\begin{align}
W_\varrho(q,p) = \frac{1}{2\pi}\int\rmd y\; \rme^{-\rmi py}\; 
   \la q+y/2|\, \varrho\, |q-y/2\ra \; .
\end{align}
This is also called the Weyl symbol of the density matrix $\varrho$. Now, the 
expectation value of any observable $\hat A$ can be calculated as a phase space 
integral:
\begin{align}
{\rm tr}[\, \hat A\, \varrho\, ] = \iint\rmd p\, \rmd q\; W_A(q,p)\; 
   W_\varrho(q,p)\; ,
\label{trarho}\end{align}
where $W_A(q,p)$ is the Weyl symbol~\cite{weyl27,Kl09} of the observable 
$\hat A$. In order to transfer the evolution equations~(\ref{me00}-\ref{me01}) 
into phase space, we need to know how multiplication with position and momentum 
operators from left and right is translated to the Wigner function 
representation. It is easily verified~\cite{Rig11}
\begin{align}
\hat q\; \varrho \; \mapsto\; \Big ( 
   q - \frac{\rmi}{2}\, \partial_p \Big )\; W_\varrho
\; , \quad
\hat p\; \varrho \; \mapsto\; \Big ( 
   p + \frac{\rmi}{2}\, \partial_q \Big )\; W_\varrho \notag\\
\varrho\;\hat q \; \mapsto\; \Big ( 
   q + \frac{\rmi}{2}\, \partial_p \Big )\; W_\varrho
\; , \quad
 \varrho\; \hat p \; \mapsto\; \Big ( 
   p - \frac{\rmi}{2}\, \partial_q \Big )\; W_\varrho\; .
\label{ApplyPosMomWig}\end{align}

\paragraph{Chord function (characteristic Wigner function)}
The chord function~\cite{Ozo98,Ozo02,Oz04} is defined as the 
Fourier transform of the Wigner function~\cite{BrePet02,Case08}
\begin{align}\label{Wtransf}
\rmw(k,s)&= \iint\rmd p\, \rmd q\; 
      \rme^{\rmi q k+\rmi s p}\, W_\varrho(q,p) \\\nonumber
&= \int\rmd q\; \rme^{\rmi q k}\, \la q+s/2 |\, \varrho\, | q-s/2\ra \; .
\end{align}
Due to this relation, Eq.~(\ref{ApplyPosMomWig}) can be readily translated into 
similar expressions for the application of position and momentum operators to 
the chord function (for later convenience, higher powers of $\hat x$ and 
$\hat p$ are included):
\begin{eqnarray}
\h{x}^n \h{p}^m\, \varrho &\mapsto& 
   \Big( \frac{s}{2} - \rmi \partial_k\Big)^{n}
   \Big( \frac{-k}{2} - \rmi\partial_s\Big)^{m}\rmw(k,s)\\
\varrho\, \h{x}^{n}\h{p}^{m} &\mapsto& 
   \Big( \frac{-s}{2} - \rmi \partial_k \Big)^n
   \Big( \frac{k}{2} - \rmi\partial_s \Big)^m\; \rmw(k,s)\\
\h{x}^{n}\,\varrho\, \h{p}^{m} &\mapsto& 
   \Big( \frac{s}{2} - \rmi \partial_k\Big)^{n}
   \Big( \frac{k}{2} - \rmi\partial_s\Big)^{m}\rmw(k,s)\\
\h{p}^{m}\,\varrho\, \h{x}^{n} &\mapsto& 
   \Big( \frac{-s}{2} - \rmi \partial_k\Big)^n
   \Big( \frac{-k}{2} - \rmi\partial_s \Big)^m\! \rmw(k,s)\, .
\end{eqnarray}
This also allows to obtain explicit expressions for the $n$-th order moments of
products of position and momentum operators, by taking the appropriate 
partial derivatives at the origin of the coordinate system:
\begin{align}
&\la \h{x}^{n}\ra = (-\rmi\, \partial_k)^n\, \rmw \big|_{k,s = 0}
\; , \quad
\la \h{p}^{n}\ra = (-\rmi\,\partial_s)^n\, \rmw \big|_{k,s = 0}\notag\\
&\frac{\la \h{x}^n\,\h{p}^m \ra + \la \h{p}^m\,\h{x}^n \ra}{2}
 = (-\rmi)^{n+m} \, \partial^n_k\, \partial^{m}_s \, \rmw \big|_{k,s = 0}\; .
\end{align}
These equations may explain the name ``characteristic Wigner function''.

In the next section, we use the properties presented here, to transfer the
evolution equations~(\ref{me00}-\ref{me01}) to partial differential equations
for the corresponding chord functions, which are then solved analytically. 
In order to compute the Uhlmann-Josza fidelity, we use a result from
Isar~\cite{Isar09}. For calculating the generalized fidelity, we 
compute the trace of $\varrho_{01}(t)$. In the chord function representations, 
this simply means that the respective chord function must be evaluated at 
$k,s = 0$. This follows from Eqs.~(\ref{trarho}) and~(\ref{Wtransf}).

\section{\label{ansols}Analytic solution}
In the chord function representation, the equations (\ref{me00}-\ref{me01}) for
the block matrices $\varrho_{ij}(t)$ defined in Eq.~(\ref{mama:defrho}) become 
the following set of partial differential equations:
 \begin{eqnarray}\label{mechft00}
 \h{L}_d\rmw_{00} &=& -\left(\rmi\, g\, s  + {\gamma_{+}\over 2} (k^2 + s^2)\right)\rmw_{00} \\\label{mechft01}
 \h{L}_{nd}\rmw_{01}&=&  -\left( \rmi \,\Delta + {\gamma_{+}\over 2} (k^2 + s^2)\right)\rmw_{01}\; ,
 \end{eqnarray}
where $\rmw_{00}$ ($\rmw_{01}$) is the chord function representation of 
$\varrho_{00}$ ($\varrho_{01}$) from Eqs.~(\ref{me00}-\ref{me01}), 
$\gamma_+ = \kappa\, (2\bar{n} + 1)$, and
\begin{align}\label{ldft}
\h{L}_d &= \partial_{\tau}  + (s + \kappa \,k )\partial_k 
   - (k-\kappa\, s)\partial_s \\\label{lndft}
\h{L}_{nd} &= \partial_{\tau}  + (s + \kappa\, k + 2\, g )\partial_k  
   - (k-\kappa\, s)\partial_s\; .
\end{align}
These differential equations can be solved analytically, using the method of
characteristics (see the appendix).
Thereby we find for $\rmw_{00}$:
\begin{align}\label{solw00}
\rmw_{00}(\vec{r},\, t) &= \rmw_{\mt{osc}}\big(\, \bR(-t)\, \vec{r}\,\big) \\
&\quad\times\exp\left( -{\rmi\over 2 } \, \vec{d}(t)\cdot\vec{r} 
   - {\alpha(t)\over 2}\, |\vec{r}\,|^2\, \right)\; ,
\notag\end{align}
where  the vector $\vec{r} = (k,s)^{T}$ collects the two independent variables 
of the chord function representation, and $\rmw_{\mt{osc}}(\, \vec{r}\, )$
represents the initial state of the oscillator in the chord function
representation.
Furthermore, we have introduced the following quantities:
\begin{align}\label{Matft}
\bR(t) &= \rme^{\kappa\, t}\begin{pmatrix}
   \cos t  & \sin t \\
   -\sin t & \cos t 
   \end{pmatrix} \; , \\
\vec{d}(t) &= \begin{pmatrix} d_1(t)\\ d_2(t)\end{pmatrix}\; , \quad
d_j(t) = 2\,g\,\int_0^t\rmd\tau\; R_{2j} (-\tau) \; , 
\label{chi}\end{align} 
where $R_{21}$ and $R_{22}$ are the respective matrix elements of $\bR(\tau)$,
and $\alpha(t) =  (\bar{n} + 1/2)\, ( 1 - \rme^{-2\kappa t}\, )$. The solution 
for $\rmw_{11}(\vec r,t)$ can be obtained from $\rmw_{00}(\vec r,t)$ by simply 
changing the sign of $g$:
\begin{align}
\rmw_{11}(\vec{r},\, t) &= \rmw_{\mt{osc}}\big(\, \bR(-t)\, \vec{r}\big) \\
&\quad\times\exp\left( {\rmi\over 2} \, \vec{d}(t)\cdot\vec{r} 
   - {\alpha(t)\over 2}\, |\vec{r}\,|^2\, \right) \; .
\notag\end{align}
From the Wigner function representations, calculated below, it can be seen
that $\pm\vec d(t)/2$ points at the position of the Gaussian state, as it 
evolves in phase space under the Hamiltonian $H_\pm$. In other words, its 
components are the expectation values of position and momentum as they evolve 
in time. 
 
As far as the initial conditions are concerned, Eq.~(\ref{mama:initstate}), we 
restrict ourselves to thermal or coherent Gaussian states for the oscillator, 
In the chord function representation; these states have the generic form:
\begin{equation}\label{ins}
 \rmw_{\mt{osc}}(\vec{r}) =  \exp\left( \rmi\,  \vec{x}_o\cdot \vec{r} 
 - {1\over 2} \vec{r}^{\,T} \bm{\sigma}_{o} \, \vec{r} \right) \; ,
\end{equation}
where the vector $\vec{x}_o = (x_o, p_o)^{T}$ contains the expectation 
values of position and momentum, and $\bm{\sigma}_o$ is the corresponding 
covariance matrix.
The uncertainty principle requires that $\det(\bm{\sigma}_o) \geq {1/4}$.

The chord function representation $\rmw_{01}$ of the non-diagonal block
can be obtained in a similar manner (see App.~\ref{ndft}). The result reads
\begin{align}
\rmw_{01}(\vec{r},t) &= \rmw_{\mt{osc}}\big(\, 
   \bR(-t) \vec{r} +  \vec{\eta}(-t) \,\big) \notag\\
&\quad\times\exp\left( -  {\alpha(t)\over 2}\, |\vec{r}\,|^2  
 - {\gamma_{+}\over 2}  \,\vec{\Gamma}(t)\,\cdot \vec{r}\, \right)\\
&\quad\times\exp\left(-\rmi \,\Delta\,t- {\gamma_{+}\over 2} \,\delta(t)\,
    \right)\; ,
\label{solw01}\end{align}
where we have introduced the following quantities: 
\begin{align}
\vec{\eta}(t) &= \frac{2\,g}{1+\kappa^2}\, 
 \left( \bR(-t) - \One \right)\, \begin{pmatrix} \kappa\\ 1\end{pmatrix}
= -\, \begin{pmatrix} d_2(t)\\ d_1(t)\end{pmatrix} \label{veta}\\
 \delta(t) &= \int_{0}^{t} \rmd t' |\vec{\eta}(t')|^2 
  = \int_0^t \rmd t'\, d(t')^2 \\
 \vec{\Gamma}( t ) &=  2 \int_0^{t}\rmd t' \, \bR^{T}(-t') \vec{\eta}(t')\; ,
\end{align} 
where $d(t)= |\vec d(t)| = |\vec\eta(t)|$. Due to this, the function 
$\delta(t)$ appearing in the solution for the non-diagonal term of the qubit is 
determined by the distance between the two Gaussians in phase space.

\subsection{Oscillator dynamics}
\begin{figure}
  \includegraphics[width=0.5\textwidth]{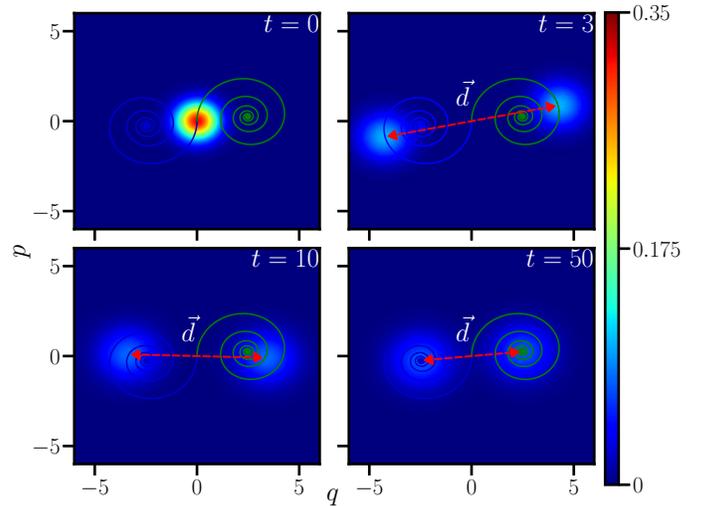}
\caption{\label{fig1}
False color plots of the Wigner function representation of the reduced 
harmonic oscillator state, for an initial product state $\varrho(0)$, 
Eq.~(\ref{mama:initstate}), built from a symmetric superposition with 
$a_{ij}= 1/2$ for the qubit and the harmonic oscillator ground state. The 
qubit-oscillator coupling is chosen as $g= 2.5$, the dissipation rate as 
$\kappa = 0.1$, and the dimensionless temperature as $D = 1$; see 
Eq.~(\ref{defnbar}). The thin solid 
lines (green and blue) show the classical trajectories under the Hamiltonians 
$H_+$ and $H_-$, respectively. The red two-sided arrow indicates the vector 
$\vec d$, introduced in Eq.(\ref{chi}). The different panels, show the Wigner
function and the vector $\vec d$ at different dimensionless times, $t=0, 3, 10,50$. }
\end{figure}

In order to illustrate the general behavior of our system, we discuss its 
reduced dynamics at very strong qubit-oscillator coupling, $g=2.5$.
As initial state we chose a product state of the form given in 
Eq.~(\ref{mama:initstate}) with $a_{ij} = 1/2$ and $\varrho_{\rm osc}$ being 
the oscillator ground state. In the present section, we consider the Wigner 
function of the reduced state after tracing over the qubit, in the next 
section~\ref{ansols:qb}, we discuss the reduced state of the qubit.

For the Wigner function of the reduced oscillator state, we find
\begin{equation}\label{wigq}
W(\vec x,t) = a_{00}\; W_+(\vec x,t) + a_{11}\; W_-(\vec x,t)\; , 
\end{equation}
where the Wigner functions $W_\pm(\vec x,t)$ have the following form:
\begin{equation}
W_\pm(\vec x ,t) = \frac{1}{2\pi \sqrt{\mt{det}\bm{\sigma}(t)}}
 \rme^{ -{1\over 2}  \left( \vec{x} - \vec{x}^{\pm}_o(t)\right)^T
 \bm{\sigma}(t)^{-1} \left( \vec{x} - \vec{x}^{\pm}_o(t)\right) } \; .
\label{Wft2}
\end{equation}
The Wigner functions conserve their Gaussian shape, while their covariance 
matrix
\begin{equation}\label{corrm}
\bm{\sigma}(t) = \alpha(t)\,\One 
   + \bR^{T}(-t)\,\bm{\sigma}_o\,\bR(-t)
\end{equation}
becomes time dependent. Note the following definitions:
$\vec{x} = (q,p)^T$ and 
$\vec{x}^{\, \pm}_o (t)=\bR^{T}(-t)\, \vec{x}_o  \pm  \vec{d}(t)/2$.

In Fig.~\ref{fig1} we plot $W(\vec x,t)$ from Eq.~(\ref{wigq}) at 
different instances in time.
The figure shows how the interaction with the qubit results in the 
splitting of the initial Gaussian wave packet into two Gaussians, following the
classical trajectories of $H_+$ and $H_-$, respectively. For large times (note
that the oscillator period is $T_{\rm osc}=2\pi$), one obtains a stationary
state, where the two Gaussian wave packets are located on the $q$-axis, each
packet in the minimum of the corresponding $g$-perturbed Hamiltonian. We will
see, that the relative vector $\vec d(t)$ between the two wave packets 
determines the fidelity measures, to be discussed below.

\subsection{\label{ansols:qb} Qubit dynamics}
The reduced state of the qubit is obtained by tracing Eq.~(\ref{mama:defrho}) 
over the oscillator degrees of freedom. This corresponds to evaluating the 
solutions given in Eqs.~(\ref{solw00}) and~(\ref{solw01}), at the origin 
$\vec{r}=0$. Since $\varrho_{00}(t)$ and  $\varrho_{11}(t)$ are valid density 
matrices for all times, the diagonal elements of the qubit state remain 
constant. In contrast to that the non-diagonal element does depend on time via
\begin{align}\label{trrho01}
&{\rm Tr}[\varrho_{01}(t)\, ] = \rmw_{\mt{osc}}\big(\,\vec{\eta}(-t\,) \,\big)
\exp\left(-\rmi \Delta t- {\gamma_{+}\over 2} \,\delta(t)\, \right)\, , \\
&\text{where}\quad
 \delta(t) =  \int_{0}^{t}\rmd t' \,  d^2(t') \; .
\notag\end{align} 

\section{\label{fidelities}Fidelity measures}
\subsection{Generalized fidelity}
Within the chord function description one can directly use Eq.~(\ref{trrho01}) 
to obtain an explicit expression for the generalized fidelity, 
Eq.~(\ref{genf}). In this way, we obtain
\begin{eqnarray}\label{fidft}
 F_{\mt{gen}}(t) &=& |\rmw_{\mt{osc}}\big(\vec{\eta}(t\,)\big)|^2\,
 \exp\left(- \gamma_{+} \,\delta(t)\,     \right)\; .
\end{eqnarray}

In the case of a general initial Gaussian state as described in 
Eq.~(\ref{ins}), the generalized fidelity takes the following form:
\begin{eqnarray}\label{fidzt2}
F_{\mt{gen}}(t)  &=&
\exp\left(- \vec{\eta}^{\,T}(t)\bm{\sigma}_o\,\vec{\eta}(t) 
-  \gamma_{+}\, \delta(t)\right)\,.
\end{eqnarray}
In the rest of the paper, we concentrate on initial thermal states, where
$\bm{\sigma}_o = M\, \One$ with $M= \bar m + 1/2$, and $x_o = p_o = 0$;
see Eq.~(\ref{ins}). In that case, the generalized fidelity becomes
\begin{align}
F_{\rm gen}(t) = \exp\big [ -M\, d(t)^2 - \gamma_+\, \delta(t)\, \big ] \; .
\label{defFgenTherm}\end{align} 
Here, $\gamma_+ = N \kappa$, with $N=(2\bar{n} + 1)$ as defined below 
Eq.~(\ref{mechft01}). The function $d(t)$ is positive, with decaying 
oscillations, which tends to a constant in the long time limit. 
Correspondingly, $\delta(t)$ is increasing monotonously, becoming approximately 
linear at sufficiently long times or when averaged over several oscillator 
periods. The expression in Eq.~(\ref{defFgenTherm}) simplifies further in the 
limit of vanishing coupling, $\kappa\to 0$, where
\begin{align}
F_{\rm gen}(t) \to \exp\big [ -8M\, g^2\, (1 - \cos t)\, \big ]\; . 
\label{unitaryFid}\end{align}
Note that for $M=1/2$ only, the expression reduces to the standard fidelity of
a pure quantum state under the perturbation of a unitary evolution, as 
described in Eq.~(\ref{mama:diagFid}). 

\begin{figure}
  \includegraphics[width=0.5\textwidth]{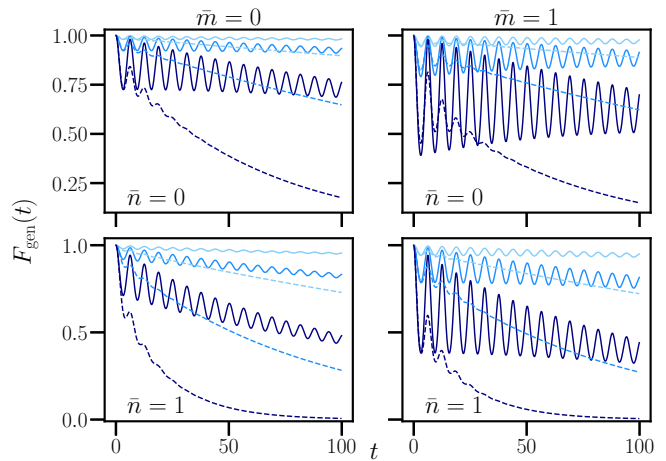}
\caption{\label{fig2} 
The generalized fidelity $F_{\rm gen}(t)$ as a function of time,
for different couplings between qubit and oscillator: $g= 0.05$ (light blue), 
$0.1$ (blue), $0.2$ (dark blue), and different environment coupling: 
$\kappa= 0.01$ (solid lines), and $0.1$ (dashed lines). The four panels show
the results for different thermal initial states (characterized by $\bar m$) 
and different temperatures of the environment (characterized by $\bar n$), as
indicated on each panel.}
\end{figure}

Figure~\ref{fig2} shows the generalized fidelity for increasing values of 
the coupling strength $g$ (from light to dark blue), different environment
couplings $\kappa$, different themperatures of the  initial states and 
the environment. The initial states are of the form given in Eq.~(\ref{ins})
with $x_o = p_o = 0$ and $\bm{\sigma}_o = (\bar{m} + 1/2)\One $, while
the temperature of the environment in characterized by the corresponding 
average number of excited modes $\bar n$, as defined in Eq.~(\ref{defnbar}).

In all cases, the generalized fidelity ultimately tends to zero in the large
time limit. Since $d(t) \to 2g/\sqrt{1+\kappa^2}$ in the limit of large times,
the slowest possible decay rate is given by 
\begin{align}
\lim_{t\to\infty} t^{-1}\; \gamma_+\, \delta(t) = \frac{\kappa}{1+\kappa^2}
   \, (2\bar n +1)\,  4g^2 \; , 
\end{align}
which can be easily calculated from Eq.~(\ref{chi}).

The exponent in Eq.~(\ref{defFgenTherm}) consists of two terms. The first term 
alone would yield decaying oscillations, in such a way that $F_{\rm gen}(t)$ 
tends to one. The initial amplitude of these oscillations is determined by $M$, 
their damping however scales with $\kappa$. The second term alone would yield a 
monotonously decaying function, with a decay rate scaling with $N \kappa g^2$.
This behavior is clearly reflected in the four panels of Fig.~\ref{fig2}, 
even though there are cases where the time range considered is too small to 
observe the complete decay.

\subsection{Uhlmann-Josza fidelity}

In Ref.~\cite{Isar09}, the author calculates the Uhlmann-Josza fidelity for
two arbitrary Gaussian states, \ie states where the position representation,
Eq.~(\ref{OpPosiRep}), of their density matrix is the exponential function of a 
quadratic polynomial in the two variables. In that case, the fidelity is 
determined completely by the first and second order moments of the position and
momentum operators. In this sense, the chord function in Eq.~(\ref{ins}) and 
the corresponding Wigner function
\begin{equation}\nonumber
W_{\rm osc}(\vec x ) = \frac{1}{ 2\pi\, \sqrt{\mt{det}(\bm{\sigma}_o)}}
 \rme^{ -{1\over 2}  \left( \vec{x} - \vec{x}_o\right)^T
 \bm{\sigma}_o^{-1} \left( \vec{x} - \vec{x}_o\right) } \; ,
\end{equation}
represent such a general Gaussian state. In that case, the first order moments
are given by $\vec x_o = (x_o,p_o)^T$, while the second order moments are 
collected in the covariance matrix $\bm{\sigma}_o$
\begin{align}
\bm\sigma_0 = \begin{pmatrix} \sigma_{11} & \sigma_{12}\\ \sigma_{12} & 
   \sigma_{22}\end{pmatrix}\; , \quad
\begin{array}{rcl}
  \sigma_{11} &=& \la\hat x^2\ra - \la\hat x\ra^2\\
  \sigma_{22} &=& \la\hat p^2\ra - \la\hat p\ra^2\\[0.5ex]
  \sigma_{12} &=& \frac{1}{2}\, 
  \la\hat x\hat p + \hat p\hat x\ra - \la\hat x\ra \la\hat p\ra \end{array}\; .
\end{align}
Assume $\varrho_1$ and $\varrho_2$ are two general Gaussian states, with
first order moments $\vec x_1$ and $\vec x_2$, as well as covariance matrices
$\bm\sigma_1$ and $\bm\sigma_2$, respectively. Then it is shown in 
Ref.~\cite{Isar09} that the Uhlmann-Josza fidelity, defined in Eq.~(\ref{uhlf}),
can be written as 
\begin{align}\label{uhls}
F_{\mt{UJ}}(\varrho_1,\varrho_2) 
 = \frac{1}{\sqrt{\mu + 4\nu} - \sqrt{4\nu}}\;
   \rme^{- \frac{1}{2}\, \vec{d}{}^T\, (\bm\sigma_1 + \bm\sigma_2)^{-1}\, 
   \vec{d}} \; ,
\end{align}
where $\mu= {\rm det}(\bm\sigma_1 + \bm\sigma_2)$, 
$\nu= [\, {\rm det}(\bm\sigma_1) - 1/4] [\, {\rm det}(\bm\sigma_2) - 1/4]$ and
$\vec d = \vec x_2 - \vec x_1$.
 
In order apply Isar's result, Eq.~(\ref{uhls}), to the density matrices 
$\varrho_{00}(t)$ and $\varrho_{11}(t)$, as prescribed in Eq.~(\ref{uhlf}), we 
note that the vector $\vec d$ must be chosen as $\vec d(t)$ from 
Eq.~(\ref{chi}), since it is exactly the distance vector between the first 
order moments of $\varrho_{00}(t)$ and $\varrho_{11}(t)$ in phase space. The
corresponding covariance matrices are the same and equal to $\bm\sigma(t)$, as
given in Eq.~(\ref{corrm}).  Therefore, we find
\[ \mu = 4\, \det[\bm\sigma(t)]\; , \quad
4\nu = \frac{[\, 4\, \det[\bm\sigma(t)] - 1\, ]^2}{4} \; , \]
which yields $\sqrt{\mu + 4\nu} =  2\, \det[\bm\sigma(t)] + 1/2$ and thereby
$\sqrt{\mu + 4\nu} - \sqrt{4\nu} = 1$. Thus, we are left with
\begin{align}\label{uhl}
F_{\mt{UJ}}(t) &= \exp\left(-\, \frac{1}{4}
\,\vec{d}^{\;T}(t)\, \bm{\sigma}^{-1}(t)\vec{d}(t)\,\right) \; , 
\end{align}
where $\bm\sigma(t)$, given in Eq.~(\ref{corrm}), simplifies to
\[ \bm\sigma(t) = \alpha(t)\, \One + \rme^{-2\kappa\, t}\, M\, \One 
   = \left[ N + (M-N)\, \rme^{-2\kappa t}\right ]\, \One \; , \]
in the case of a thermal initial state with $\bm\sigma_0 = M\, \One$. In that
latter case,
\begin{align}
F_{\mt{UJ}}(t) &= \exp\left(-\, \frac{1}{4}\, 
   \frac{d(t)^2}{N + (M-N)\, \rme^{-2\kappa t}}\, \right)\; .
\label{fUJthermal}\end{align}

\begin{figure}
\includegraphics[width=0.5\textwidth]{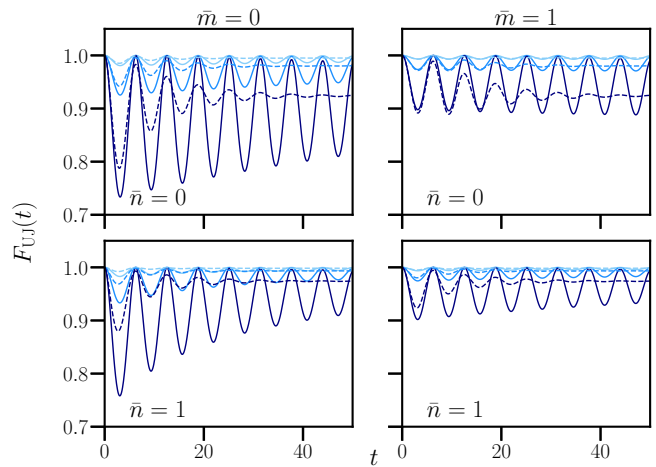}
\caption{\label{fig3} 
The Uhlmann-Josza fidelity $F_{\rm UJ}(t)$ as a function of time, for 
different couplings between qubit and oscillator: $g=0.05$ (light blue), $0.1$ 
(blue), $0.2$ (dark blue), and different environment coupling $\kappa= 0.01$
(solid lines), and $0.1$ (dashed lines). The four panels show the results for
different thermal initial states (characterized by $\bar m$) and different 
temperatures of the environment (characterized by $\bar n$), as indicated on 
each panel.} 
\end{figure}

Figure~\ref{fig3} shows the time evolution of the Uhlmann-Josza fidelity
in the same conditions and with the same line types and color codings as in 
the case of the generalized fidelity shown in Fig.~\ref{fig2}. 
The most striking difference between the two fidelities can be observed in 
their behavior at long times. While the generalized fidelity ultimately decays 
to zero in all case, the Uhlmann-Josza fidelity tends to the constant
\begin{align}
\lim_{t\to\infty} F_{\rm UJ}(t) = \exp\left( \frac{-g^2}{N\, (1+\kappa^2)}
   \right ) 
\end{align} 
[see Eq.~(\ref{d2fullanalytic}), below]. Both quantities strongly depend 
on $d(t)$, and by consequence show very similar oscillatory behavior with 
a period similar to the fundamental oscillator period.

\subsection{\label{connections}Connecting quantities}

In this section, we discuss the possibility to use the qubit as a probe 
system for extracting information about the evolution of the oscillator in 
contact with a heat bath. Evidently, all the information extractable from the
qubit must be contained in the generalized fidelity $F_{\rm gen}(t)$. We limit
ourselves to initial thermal states of the oscillator -- not necessarily in
equilibrium with the heat bath.\\

(i) In a first approach, we simply take advantage of the fact that the 
behavior of the system is known analytically. In principle, it is therefore
sufficient to determine all relevant parameters of the system in order to 
determine its dynamics. In our particular case, these parameters are: $g$ the 
coupling between qubit and oscillator, $\kappa$ the coupling between the 
oscillator and the heat bath, and finally $M$ and $N$ which characterize the
temperature of the initial state and the heat bath respectively. Calculating
the logarithmic derivative of the generalized fidelity, we find the following
analytic expression:
\begin{align}
H(t) &= -\, \frac{\rmd}{\rmd t}\, \ln\big [ F_{\rm gen}(t)\, \big ]
      = M\, \frac{\rmd}{\rmd t}\, d^2(t) + 2\kappa\, N\; d^2(t) \; , \notag\\
d^2(t) &=  {4\, g^2\over 1 + \kappa^2}
      \left(\rme^{-2\, \kappa\, t} - 2\rme^{-\kappa\, t}\cos t + 1\right)\; .
\label{d2fullanalytic}\end{align}
Thus, in principle, it seems that the function $H(t)$ depends on all four 
parameters in an independent way. Therefore, a non-linear parameter fit may
be used to estimate their values.\\

(ii) As an alternative, we could try to determine the function $d^2(t)$
directly, using the fact that it only depends on $\kappa$ and $g$ but not
its particular form. This is possible by measuring $F_{\rm gen}(t)$ for two
initial states of different temperatures, $M_1$ and $M_2$ (assuming these are
known a priori). In that case, one
obtains:
\begin{align}
\frac{\ln[ F_{\rm gen}^{(M_1)}(t) ] - \ln[ F_{\rm gen}^{(M_2)}(t)]}
   {M_2 - M_1} = d^2(t) \; .
\end{align}
Now, we can estimate $g$ and $\kappa$ separately from the behavior of $d^2(t)$.
Then, in a second step, we estimate the temperature $N$ of the heat bath from 
the identity
\begin{align}
\frac{M_2\, \ln[ F_{\rm gen}^{(M_1)}(t) ] - M_1\, \ln[ F_{\rm gen}^{(M_2)}(t)]}
   {M_1 - M_2} = \kappa\, N\; \delta(t) \; .
\end{align}
This method may be more robust as the first one, since we do not estimate so
many parameters from one single function.\\ 

\paragraph*{Uhlmann-Josza fidelity}
Once, the function $d^2(t)$ is known, together with the parameters $N,M$ and
$\kappa$, we can reconstruct the Uhlmann-Josza fidelity with the help of 
Eq.~(\ref{fUJthermal}). At the moment, it is still an open question, whether 
this or a similar relation may hold in more general cases also. These cases
may include: 1. different initial states for instance coherent states away from
the equilibrium point, 2. cat states -- i.e. superpositions of coherent 
states, etc.\\

\paragraph*{Purities of the reduced states} As it turns out, the purities of the
reduced states, of the qubit but also of the oscillator, can be related in a
very similar manner to the respective quantum fidelities. In the case of the 
qubit reduced state, this is fairly obvious:
\begin{equation}\label{pq}
P_{\mt{q}}(t) =  a^2_{00} + a^2_{11} 
   + 2\, |a_{01}|^2 \, F^{(M)}_{\mt{gen}}(t)\; .
\end{equation}
In the case of the oscillator reduced state, this follows from the fact that
\[ P_{\mt{osc}}(t) = 2\pi\iint\rmd p\, \rmd q \; W^2(\vec{x},t)\; , \]
with $W(\vec{x},t)$ given by (\ref{wigq}). In that case,
\begin{align}
P_{\mt{osc}}(t)= \frac{a^2_{00} + a^2_{11} + 2\, a_{00}\,a_{11} \, 
   F^{(M)}_{\mt{UJ}}(t)}{2\, \sqrt{\det\bm{\sigma}(t)}} \; .
\label{posc}\end{align}

\section{\label{con}Conclusions}
We considered an open quantum system consisting of a harmonic oscillator being 
coupled to a finite-temperature heat bath, equipped with an additional two-level
system (qubit). This qubit is coupled to the oscillator via dephasing coupling,
and serves as a probe for the dynamics of the system. The quantum master 
equation, which describes the evolution of the whole system, can be solved 
analytically, using the Fourier transform of the Wigner function, the so called 
``chord function''.

Provided the two-level probe is initially prepared in a superposition state, 
the loss of coherence over time provides sufficient information in order to
determine the complete dynamics of the dissipative oscillator. In other words,
based on the decoherence function, we can estimate all relevant parameters of
the dynamics: the temperature of the heat bath, that of the initial state 
(\eg{} in the case we are interested in a temperature quench), the coupling 
strength to the bath as well as that of the probe. 

The present setup, provides the rare opportunity to study analytically the 
behavior of the Uhlmann-Josza fidelity between mixed states subject to 
different evolutions in time. We use this to investigate similarities and 
differences between the decoherence function, which is almost identical to 
the generalized fidelity as introduced in Ref.~\cite{GMS16RTSA}, and the 
standard Uhlmann-Josza fidelity. 

So far, we restricted ourselves to thermal initial states. In that case, the
coupling to the qubit gives rise to two different evolutions governed by
separated harmonic potentials. It would be interesting to study more general
initial states, such as displaced Gaussian states, similar to those which 
have been analyzed in the Jaynes-Cummings model of cavity QED~\cite{PurAga87},
or cat states; or one could even consider Schr\" odinger cat states as initial 
states. That would allow us to investigate the relation between generalized and 
Uhlmann-Josza fidelity from a more general perspective.

In addition, we may find interesting applications in the area of quantum 
thermodynamics. For instance, we may realize Carnot cycles with the harmonic
oscillator at finite times, and monitor the evolving state with the help of 
the coupled qubit. Since the coupling between qubit and oscillator is of 
dephasing type, the systems cannot interchange energy thus one can observe the
evolution of the thermodynamic system without affecting its thermodynamic
properties. A possible experimental realization could be build from two-level
atoms in an harmonic trap, where the dephasing coupling and the measurement of
the decoherence function is easy to achieve~\cite{GCZ97,GPSS04,Haug05}.

\appendix
\section{\label{app1} Derivation of the solutions of the generalized master 
   equations}

By using  the notation $\langle i|\varrho|j\rangle = \varrho_{ij}$ for 
$i,j=0,1$ represent the projection of the system into the different states of 
the qubit. The matrix elements master equations may be written as:
\begin{eqnarray}\label{ft00}
\rmi\, \frac{\rmd\varrho_{00}}{\rmd t} &=&
  [H_{\mt{osc}},\varrho_{00}] + g\,[\h{x},\varrho_{00}] +\rmi \mathcal{L}[\varrho_{00}]\\\label{ft01}
\rmi\, \frac{\rmd\varrho_{01}}{\rmd t} &=&
  [ H_{\mt{osc}},\varrho_{01}] + \{ \Delta/2 +  g\, \h{x},\varrho_{01}\} + \rmi\mathcal{L}[\varrho_{01}]
\end{eqnarray}
where $\mathcal{L}[\varrho_{ij}]$ is given by (\ref{Lindo}). The solution to 
the matrix elements master equations can be more easily carried out by
employing the chord function description. By doing the transformations, matrix 
elements master equations can be written as a set of partial differential 
equations:
 \begin{eqnarray}\label{mechft00ap}
 \h{L}_d\rmw_{00} &=& -\left(\rmi\, g\, s  + {\gamma_{+}\over 2} (k^2 + s^2)\right)\rmw_{00} \\\label{mechft01ap}
 \h{L}_{nd}\rmw_{01}&=&  -\left( \rmi \,\Delta + {\gamma_{+}\over 2} (k^2 + s^2)\right)\rmw_{01}
 \end{eqnarray}
where  we have defined  $ \gamma_{+} = 2\kappa (\bar{n} + 1/2)$, and 
\begin{eqnarray}\label{ldftap}
 \h{L}_d &=& \partial_{t}  + (s + \kappa \,k )\partial_k - (k-\kappa\, s)\partial_s \\\label{lndftap}
 \h{L}_{nd} &=& \partial_{t}  + (s + \kappa\, k + 2\, g )\partial_k  - (k-\kappa\, s)\partial_s
 \end{eqnarray}

\subsection{\label{dft}Diagonal element $\rmw_{00}$}

For equation~(\ref{mechft00ap}) one can write down the set of parametric 
differential equations in the following form:
\begin{align}\label{peft1d}
\frac{\rmd k}{\rmd t} &= s +\kappa\, k \; , \quad 
\frac{\rmd s}{\rmd t} = - k + \kappa\, s \\
\frac{\rmd\rmw_{00}}{\rmd t} &= -\left[\, \rmi\, g s + \frac{\gamma_+}{2}\,
   (k^2 +s^2)\, \right]\, \rmw_{00}\; ,
\label{peft2d}\end{align}
and by coupling the first two equations (\ref{peft1d}), we can write down a 
second order ordinary differential equation for $k$:
\begin{equation}\label{2ndok}
 \ddot{k} -\beta \dot{k} + \omega^2 k = 0,
\end{equation}
where $\omega^2 = 1 + \kappa^2$. The solution of equation
(\ref{2ndok}) may be written as:
 \begin{equation}\label{sk}
k(\tau) = \rme^{\kappa \, t}\, \big ( a_1\, \sin t + a_2\,
               \cos t\big )
\end{equation}
where $a_1$ and $a_2$ are the characteristic curves which remain constant for all time.
The variable $s(t)$ may be obtained through first equation of (\ref{peft1d}),
$s = \dot{k} - \kappa\, k$ yielding:
\begin{equation}
 s( t) = \rme^{\kappa t }\, \big (  a_1\, \cos t - a_2\,\sin t \big )\,. 
\end{equation}
For these type of linear differential equations, one can always 
define the fundamental matrix which maps any point 
$(k(t'),s(t'))$ at the time $t'$, 
along the characteristics to any other point $(k(t),s(t))$
at time $t$  as  $\vec{r}(t) = \bR(t-t')\vec{r}(t')$. For
time invariant systems where the parameter are no time dependent, 
the fundamental matrix has a closed form. In the present case, 
this one has  the following form:
\begin{equation}\label{Matfta}
\bR(t) = \rme^{\kappa t }\begin{pmatrix}
   \cos t  & \sin t \\
   -\sin t & \cos t 
   \end{pmatrix}
\; .
\end{equation}
This fundamental matrix  has the property
that it only depends of the difference of the initial and final time 
and posses group properties in the sense that
$\bR(t_2-t_0) = \bR(t_2-t_1)\bR(t_1-t_0)$ for all times except 
in the limit where $t\rightarrow \pm\infty$ for which it becomes singular. 
At any other time, the fundamental matrix always fulfills
$\bR(-t) = \bR^{-1}(t)$. The integration of (\ref{mechft00ap}) 
is done as follows; 
\begin{eqnarray}\label{intft3d}
\int_{\rmw_{00}(0)}^{\rmw_{00}(t)}{\rmd\rmw_{00}\over \rmw_{00}}&=& -
        \rmi\, g\,\int_{0}^{t}\rmd t'\, s(t') \\\nonumber
        && -{\gamma_{+}\over 2}\int_{0}^{t}\rmd  t' \,   \left( k^2(t') +  s^2(t')\right)\; .
\end{eqnarray}
 Within the fundamental matrix, the
 integration over the right hand side 
 can be calculated by using the fact that
 \begin{equation} 
 \vec{r}\,(t') = \bR(t'-t)\,\vec{r}(t) \; ,
 \end{equation}
thus, we can write down an explicitly expression for the 
evolution of this chord function matrix element as:
\begin{eqnarray}\label{solw00a}
\rmw_{00}(\vec{r},t)& =& 
\rmw_{00}\big(\, \bR(-t) \vec{r}\, ,\, \tau \big)\\\nonumber
&&\exp\left( -{\rmi\over 2 } \vec{d}(t)\cdot\vec{r} - {\gamma_{+}\over 2}\,\alpha(t)\, |\vec{r}\,|^2\, \right)
\end{eqnarray}
where $\alpha(t)$ is given by 
\begin{align}
\alpha(t)= \int_0^t\rmd t'\; \left[ \, R^2_{11}(-t') + R^2_{12}(-t')\, \right]
  = {1-\rme^{-\kappa t}\over \kappa}\, ,
\end{align}
and $\vec{d}(t)= (d_1(t),d_2(t))^T$ where its components have the following 
form ($j= 1,2$):
\begin{align}\label{chia}
d_j(t) &= 2\, g\, \int_0^t\rmd t'\; R_{2j} (-t')\; .
\end{align}
where $R_{21}$ and $R_{22}$ are matrix elements of the map described in 
Eq.~(\ref{Matfta}).

\subsection{\label{ndft}Non-diagonal element $\rmw_{01}$}

For the non-diagonal element master equation (\ref{mechft01ap}), its 
parametric form  is given by:
\begin{align}\label{peft1}
\frac{\rmd k}{\rmd t} &= s +\kappa k + 2g\; , \quad 
\frac{\rmd s}{\rmd t} = - k + \kappa s \; , \\
\frac{\rmd\rmw_{01}}{\rmd t} &= - \left(  \rmi \Delta + \frac{\gamma_+}{2}
  \, (k^2 +s^2)\right)\, \rmw_{01}\; .
\label{peft2}\end{align}
As before, one can write down a second order ordinary differential equation for
$k$ by coupling the first two equations (\ref{peft1}) yielding:
\begin{equation}\label{2ndoknd}
 \ddot{k} -2\,\kappa \dot{k} + (1 + \kappa^2)\, k = -2\,\kappa\, g,
\end{equation}
The solution of equation (\ref{2ndoknd}) may be written as:
 \begin{equation}\label{sknd}
k(t) = \rme^{\kappa t}\, \big ( a_1\, \sin t + a_2\,
               \cos t \big )  -2\,g\, \kappa_1,
\end{equation}
where $\kappa_1 = \kappa /(1+\kappa^2)$. By solving for $s$ in the first parametric equation 
and plugging in, the result for $k$ one has for $s$:
\begin{eqnarray}\label{ss}
 s( t ) &= & \dot{k} - \gamma k - 2\,g\\\nonumber
 &=& \rme^{\kappa t}\, \big ( - a_1\, \cos t + a_2\,
               \sin t\big )  - 2\,g\,\kappa_2\;.
\end{eqnarray}
where $\kappa_2 = \kappa_1 /\kappa $.
Thus, by doing the following change of variables;
\begin{eqnarray}
 k'(t) = k(t) - 2\,g\,\kappa_1 , \\
 s'(t) = s(t) - 2\,g \,\kappa_2
\end{eqnarray}
then, we can describe the map given by the matrix $\bR$
given at (\ref{Matfta}) to the primed variables exactly
as we did for the diagonal element 
case, \ie: $\vec{r}\,'(t) = \bR(t-t') \vec{r}\,'(t')$ where 
$\vec{r }\,'(t) = \vec{r}(t) + 2\,g\, \vec{\kappa}$ and
\begin{equation}\label{vgamma}
\vec{\kappa} = \begin{pmatrix}
    \kappa_1 \\
    \kappa_2
   \end{pmatrix}
\; = {1\over 1+\kappa^2} \, \begin{pmatrix}
    \kappa \\
    1
   \end{pmatrix} \,.
\end{equation}
With these redefinitions, one can now write down the
map which describes the motion of 
any point $(k(t),s(t))$ along 
the characteristics as:
\begin{equation}\label{mapftnd}
 \vec{r}(t) = \bR(t-t')\; \vec{r}(t') + 2\, g\,(\bR(t-t') - \One)\,\vec{\kappa}.
 \end{equation}
Integration of the third equation will yield
 \begin{eqnarray}\nonumber
\int_{\rmw_{01}(0)}^{\rmw_{01}(t)}{\rmd\rmw_{01}\over \rmw_{01}}&=& 
        -\int_{0}^{t}\rmd t'\,\left( {\rmi \Delta} + {\gamma_{+}\over 2} \,|\vec{r}\,( t')|^2\,\right)\; ,\\\nonumber
        &=& -\rmi \Delta t - {\gamma_{+}\over 2}\, \int_{0}^{ t }\rmd  t'\, |\vec{r}\,( t')|^2\,.\\ \label{intft3nd}
\end{eqnarray}
where $|\vec{r}(t')|^2 = k^2(t')  +  s^2(t')$. Now, 
we can use the map defined at (\ref{mapftnd}) to write
$k(\tau ')$ and $s(\tau ')$ appearing 
in the integrand of the right hand side of (\ref{intft3nd}), as:
\begin{equation}
|\vec{r}\,(t')|^2 = |\bR(t' - t)\; \vec{r}(t) + \vec{\eta}(t-t')\,|^2\; , 
\label{rtauft} \end{equation}
where
\begin{equation}
\vec\eta(t) = \frac{2\,g}{1+\kappa^2}\, 
 \left( \bR(-t) - \One \right)\, \begin{pmatrix} \kappa\\ 1\end{pmatrix}
= -\, \begin{pmatrix} d_2(t)\\ d_1(t)\end{pmatrix}\; .
\label{ap:etadef}\end{equation} 
The last equality in this equation follows from direct evaluation of the 
integrals defined in Eq.~(\ref{chia}). 
Thus, by doing some algebra the integral in (\ref{intft3nd}) can be written as: 
\begin{align}
&\int_0^{t}\rmd t' \,|\vec{r}(t')|^2 = \int_0^{t}\rmd t' \,\bigg( 
   |\bR(-t') \vec{r}(t)|^2 \notag\\
&\qquad\qquad + 2 \, \bR^{T}(-t') \vec{\eta}(t') \cdot \vec{r}(t)
   + |\vec\eta(t')|^2\, \bigg)\; .
\label{ftint}\end{align}
Integration can be easily performed yielding for the non diagonal matrix element
the following:
\begin{eqnarray}\nonumber
\rmw_{01}(\vec{r},t)& =& 
\rmw_{01}\big(\, \bR(-t) \vec{r} + \vec{\eta}(t) 
\, \big)\,\rme^{ - \rmi \Delta \,t }\\\nonumber
&&\exp\left[ - {\gamma_{+}\over 2}\, \alpha(t)\, |\vec{r}\,|^2  
- {\gamma_{+}\over 2} \delta(t)     \right]\\\label{solw01a}
&&\exp\left[ - {\gamma_{+}\over 2} \, \vec{\Gamma}( t )\,\cdot \vec{r}    
\,     \right]\,,
\end{eqnarray}
where,
\begin{align}
 \delta(t) &= \int_{0}^{t} \rmd t' |\vec{\eta}(t')|^2\; ,  \\
 \vec\Gamma(t) &=  2 \int_0^{t}\rmd t' \, \bR^{T}(-t')\, \vec\eta(t')\; .
\end{align}

\bibliographystyle{apsrev4-1}
\bibliography{references}

\begin{thebibliography}{60}%
\makeatletter
\providecommand \@ifxundefined [1]{%
 \@ifx{#1\undefined}
}%
\providecommand \@ifnum [1]{%
 \ifnum #1\expandafter \@firstoftwo
 \else \expandafter \@secondoftwo
 \fi
}%
\providecommand \@ifx [1]{%
 \ifx #1\expandafter \@firstoftwo
 \else \expandafter \@secondoftwo
 \fi
}%
\providecommand \natexlab [1]{#1}%
\providecommand \enquote  [1]{``#1''}%
\providecommand \bibnamefont  [1]{#1}%
\providecommand \bibfnamefont [1]{#1}%
\providecommand \citenamefont [1]{#1}%
\providecommand \href@noop [0]{\@secondoftwo}%
\providecommand \href [0]{\begingroup \@sanitize@url \@href}%
\providecommand \@href[1]{\@@startlink{#1}\@@href}%
\providecommand \@@href[1]{\endgroup#1\@@endlink}%
\providecommand \@sanitize@url [0]{\catcode `\\12\catcode `\$12\catcode
  `\&12\catcode `\#12\catcode `\^12\catcode `\_12\catcode `\%12\relax}%
\providecommand \@@startlink[1]{}%
\providecommand \@@endlink[0]{}%
\providecommand \url  [0]{\begingroup\@sanitize@url \@url }%
\providecommand \@url [1]{\endgroup\@href {#1}{\urlprefix }}%
\providecommand \urlprefix  [0]{URL }%
\providecommand \Eprint [0]{\href }%
\providecommand \doibase [0]{http://dx.doi.org/}%
\providecommand \selectlanguage [0]{\@gobble}%
\providecommand \bibinfo  [0]{\@secondoftwo}%
\providecommand \bibfield  [0]{\@secondoftwo}%
\providecommand \translation [1]{[#1]}%
\providecommand \BibitemOpen [0]{}%
\providecommand \bibitemStop [0]{}%
\providecommand \bibitemNoStop [0]{.\EOS\space}%
\providecommand \EOS [0]{\spacefactor3000\relax}%
\providecommand \BibitemShut  [1]{\csname bibitem#1\endcsname}%
\let\auto@bib@innerbib\@empty
\bibitem [{\citenamefont {Gardiner}\ \emph {et~al.}(1997)\citenamefont
  {Gardiner}, \citenamefont {Cirac},\ and\ \citenamefont {Zoller}}]{GCZ97}%
  \BibitemOpen
  \bibfield  {author} {\bibinfo {author} {\bibfnamefont {S.~A.}\ \bibnamefont
  {Gardiner}}, \bibinfo {author} {\bibfnamefont {J.~I.}\ \bibnamefont {Cirac}},
  \ and\ \bibinfo {author} {\bibfnamefont {P.}~\bibnamefont {Zoller}},\
  }\href@noop {} {\bibfield  {journal} {\bibinfo  {journal} {Phys. Rev. Lett.}\
  }\textbf {\bibinfo {volume} {79}},\ \bibinfo {pages} {4790} (\bibinfo {year}
  {1997})}\BibitemShut {NoStop}%
\bibitem [{\citenamefont {Peres}(1984)}]{Peres84}%
  \BibitemOpen
  \bibfield  {author} {\bibinfo {author} {\bibfnamefont {A.}~\bibnamefont
  {Peres}},\ }\href@noop {} {\bibfield  {journal} {\bibinfo  {journal} {Phys.
  Rev. A}\ }\textbf {\bibinfo {volume} {30}},\ \bibinfo {pages} {1610}
  (\bibinfo {year} {1984})}\BibitemShut {NoStop}%
\bibitem [{\citenamefont {Pastawski}\ \emph {et~al.}(2000)\citenamefont
  {Pastawski}, \citenamefont {Levstein}, \citenamefont {Usaj}, \citenamefont
  {Raya},\ and\ \citenamefont {Hirschinger}}]{Pasta00}%
  \BibitemOpen
  \bibfield  {author} {\bibinfo {author} {\bibfnamefont {H.~M.}\ \bibnamefont
  {Pastawski}}, \bibinfo {author} {\bibfnamefont {P.~R.}\ \bibnamefont
  {Levstein}}, \bibinfo {author} {\bibfnamefont {G.}~\bibnamefont {Usaj}},
  \bibinfo {author} {\bibfnamefont {J.}~\bibnamefont {Raya}}, \ and\ \bibinfo
  {author} {\bibfnamefont {J.}~\bibnamefont {Hirschinger}},\ }\href@noop {}
  {\bibfield  {journal} {\bibinfo  {journal} {Physica (Amsterdam) A}\ }\textbf
  {\bibinfo {volume} {283}},\ \bibinfo {pages} {166} (\bibinfo {year}
  {2000})}\BibitemShut {NoStop}%
\bibitem [{\citenamefont {Gorin}\ \emph {et~al.}(2006)\citenamefont {Gorin},
  \citenamefont {Prosen}, \citenamefont {Seligman},\ and\ \citenamefont
  {\v{Z}nidari\v{c}}}]{GPSZ06}%
  \BibitemOpen
  \bibfield  {author} {\bibinfo {author} {\bibfnamefont {T.}~\bibnamefont
  {Gorin}}, \bibinfo {author} {\bibfnamefont {T.}~\bibnamefont {Prosen}},
  \bibinfo {author} {\bibfnamefont {T.~H.}\ \bibnamefont {Seligman}}, \ and\
  \bibinfo {author} {\bibfnamefont {M.}~\bibnamefont {\v{Z}nidari\v{c}}},\
  }\href@noop {} {\bibfield  {journal} {\bibinfo  {journal} {Phys. Rep.}\
  }\textbf {\bibinfo {volume} {435}},\ \bibinfo {pages} {33} (\bibinfo {year}
  {2006})}\BibitemShut {NoStop}%
\bibitem [{\citenamefont {Jacquod}\ and\ \citenamefont
  {Petitjean}(2009)}]{JacPet09}%
  \BibitemOpen
  \bibfield  {author} {\bibinfo {author} {\bibfnamefont {P.}~\bibnamefont
  {Jacquod}}\ and\ \bibinfo {author} {\bibfnamefont {C.}~\bibnamefont
  {Petitjean}},\ }\href {\doibase 10.1080/00018730902831009} {\bibfield
  {journal} {\bibinfo  {journal} {Advances in Physics}\ }\textbf {\bibinfo
  {volume} {58}},\ \bibinfo {pages} {67} (\bibinfo {year} {2009})}\BibitemShut
  {NoStop}%
\bibitem [{\citenamefont {Binder}\ \emph {et~al.}(2015)\citenamefont {Binder},
  \citenamefont {Vinjanampathy}, \citenamefont {Modi},\ and\ \citenamefont
  {Goold}}]{Bi15}%
  \BibitemOpen
  \bibfield  {author} {\bibinfo {author} {\bibfnamefont {F.}~\bibnamefont
  {Binder}}, \bibinfo {author} {\bibfnamefont {S.}~\bibnamefont
  {Vinjanampathy}}, \bibinfo {author} {\bibfnamefont {K.}~\bibnamefont {Modi}},
  \ and\ \bibinfo {author} {\bibfnamefont {J.}~\bibnamefont {Goold}},\ }\href
  {\doibase 10.1103/PhysRevE.91.032119} {\bibfield  {journal} {\bibinfo
  {journal} {Phys. Rev. E}\ }\textbf {\bibinfo {volume} {91}},\ \bibinfo
  {pages} {032119} (\bibinfo {year} {2015})}\BibitemShut {NoStop}%
\bibitem [{\citenamefont {Ng}\ \emph {et~al.}(2015)\citenamefont {Ng},
  \citenamefont {Mančinska}, \citenamefont {Cirstoiu}, \citenamefont
  {Eisert},\ and\ \citenamefont {Wehner}}]{Ng15}%
  \BibitemOpen
  \bibfield  {author} {\bibinfo {author} {\bibfnamefont {N.~H.~Y.}\
  \bibnamefont {Ng}}, \bibinfo {author} {\bibfnamefont {L.}~\bibnamefont
  {Mančinska}}, \bibinfo {author} {\bibfnamefont {C.}~\bibnamefont
  {Cirstoiu}}, \bibinfo {author} {\bibfnamefont {J.}~\bibnamefont {Eisert}}, \
  and\ \bibinfo {author} {\bibfnamefont {S.}~\bibnamefont {Wehner}},\ }\href
  {http://stacks.iop.org/1367-2630/17/i=8/a=085004} {\bibfield  {journal}
  {\bibinfo  {journal} {New Journal of Physics}\ }\textbf {\bibinfo {volume}
  {17}},\ \bibinfo {pages} {085004} (\bibinfo {year} {2015})}\BibitemShut
  {NoStop}%
\bibitem [{\citenamefont {Millen}\ and\ \citenamefont {Xuereb}(2016)}]{Mi16}%
  \BibitemOpen
  \bibfield  {author} {\bibinfo {author} {\bibfnamefont {J.}~\bibnamefont
  {Millen}}\ and\ \bibinfo {author} {\bibfnamefont {A.}~\bibnamefont
  {Xuereb}},\ }\href {http://stacks.iop.org/1367-2630/18/i=1/a=011002}
  {\bibfield  {journal} {\bibinfo  {journal} {New Journal of Physics}\ }\textbf
  {\bibinfo {volume} {18}},\ \bibinfo {pages} {011002} (\bibinfo {year}
  {2016})}\BibitemShut {NoStop}%
\bibitem [{\citenamefont {Goold}\ \emph {et~al.}(2016)\citenamefont {Goold},
  \citenamefont {Huber}, \citenamefont {Riera}, \citenamefont {del Rio},\ and\
  \citenamefont {Skrzypczyk}}]{Go16}%
  \BibitemOpen
  \bibfield  {author} {\bibinfo {author} {\bibfnamefont {J.}~\bibnamefont
  {Goold}}, \bibinfo {author} {\bibfnamefont {M.}~\bibnamefont {Huber}},
  \bibinfo {author} {\bibfnamefont {A.}~\bibnamefont {Riera}}, \bibinfo
  {author} {\bibfnamefont {L.}~\bibnamefont {del Rio}}, \ and\ \bibinfo
  {author} {\bibfnamefont {P.}~\bibnamefont {Skrzypczyk}},\ }\href
  {http://stacks.iop.org/1751-8121/49/i=14/a=143001} {\bibfield  {journal}
  {\bibinfo  {journal} {J Phys A Math Theor}\
  }\textbf {\bibinfo {volume} {49}},\ \bibinfo {pages} {143001} (\bibinfo
  {year} {2016})}\BibitemShut {NoStop}%
\bibitem [{\citenamefont {Jevtic}\ \emph {et~al.}(2015)\citenamefont {Jevtic},
  \citenamefont {Newman}, \citenamefont {Rudolph},\ and\ \citenamefont
  {Stace}}]{Jev15}%
  \BibitemOpen
  \bibfield  {author} {\bibinfo {author} {\bibfnamefont {S.}~\bibnamefont
  {Jevtic}}, \bibinfo {author} {\bibfnamefont {D.}~\bibnamefont {Newman}},
  \bibinfo {author} {\bibfnamefont {T.}~\bibnamefont {Rudolph}}, \ and\
  \bibinfo {author} {\bibfnamefont {T.~M.}\ \bibnamefont {Stace}},\ }\href
  {\doibase 10.1103/PhysRevA.91.012331} {\bibfield  {journal} {\bibinfo
  {journal} {Phys. Rev. A}\ }\textbf {\bibinfo {volume} {91}},\ \bibinfo
  {pages} {012331} (\bibinfo {year} {2015})}\BibitemShut {NoStop}%
\bibitem [{\citenamefont {Mancino}\ \emph {et~al.}(2017)\citenamefont
  {Mancino}, \citenamefont {Sbroscia}, \citenamefont {Gianani}, \citenamefont
  {Roccia},\ and\ \citenamefont {Barbieri}}]{Man17}%
  \BibitemOpen
  \bibfield  {author} {\bibinfo {author} {\bibfnamefont {L.}~\bibnamefont
  {Mancino}}, \bibinfo {author} {\bibfnamefont {M.}~\bibnamefont {Sbroscia}},
  \bibinfo {author} {\bibfnamefont {I.}~\bibnamefont {Gianani}}, \bibinfo
  {author} {\bibfnamefont {E.}~\bibnamefont {Roccia}}, \ and\ \bibinfo {author}
  {\bibfnamefont {M.}~\bibnamefont {Barbieri}},\ }\href {\doibase
  10.1103/PhysRevLett.118.130502} {\bibfield  {journal} {\bibinfo  {journal}
  {Phys. Rev. Lett.}\ }\textbf {\bibinfo {volume} {118}},\ \bibinfo {pages}
  {130502} (\bibinfo {year} {2017})}\BibitemShut {NoStop}%
\bibitem [{\citenamefont {Quan}\ \emph {et~al.}(2007)\citenamefont {Quan},
  \citenamefont {Liu}, \citenamefont {Sun},\ and\ \citenamefont
  {Nori}}]{Qua07}%
  \BibitemOpen
  \bibfield  {author} {\bibinfo {author} {\bibfnamefont {H.~T.}\ \bibnamefont
  {Quan}}, \bibinfo {author} {\bibfnamefont {Y.-x.}\ \bibnamefont {Liu}},
  \bibinfo {author} {\bibfnamefont {C.~P.}\ \bibnamefont {Sun}}, \ and\
  \bibinfo {author} {\bibfnamefont {F.}~\bibnamefont {Nori}},\ }\href {\doibase
  10.1103/PhysRevE.76.031105} {\bibfield  {journal} {\bibinfo  {journal} {Phys.
  Rev. E}\ }\textbf {\bibinfo {volume} {76}},\ \bibinfo {pages} {031105}
  (\bibinfo {year} {2007})}\BibitemShut {NoStop}%
\bibitem [{\citenamefont {Uzdin}\ \emph {et~al.}(2015)\citenamefont {Uzdin},
  \citenamefont {Levy},\ and\ \citenamefont {Kosloff}}]{Ra15}%
  \BibitemOpen
  \bibfield  {author} {\bibinfo {author} {\bibfnamefont {R.}~\bibnamefont
  {Uzdin}}, \bibinfo {author} {\bibfnamefont {A.}~\bibnamefont {Levy}}, \ and\
  \bibinfo {author} {\bibfnamefont {R.}~\bibnamefont {Kosloff}},\ }\href
  {\doibase 10.1103/PhysRevX.5.031044} {\bibfield  {journal} {\bibinfo
  {journal} {Phys. Rev. X}\ }\textbf {\bibinfo {volume} {5}},\ \bibinfo {pages}
  {031044} (\bibinfo {year} {2015})}\BibitemShut {NoStop}%
\bibitem [{\citenamefont {Horodecki}\ and\ \citenamefont
  {Oppenheim}(2013)}]{Ho13}%
  \BibitemOpen
  \bibfield  {author} {\bibinfo {author} {\bibfnamefont {M.}~\bibnamefont
  {Horodecki}}\ and\ \bibinfo {author} {\bibfnamefont {J.}~\bibnamefont
  {Oppenheim}},\ }\href {http://dx.doi.org/10.1038/ncomms3059} {\bibfield
  {journal} {\bibinfo  {journal} {Nature Communications}\ }\textbf {\bibinfo
  {volume} {4}},\ \bibinfo {pages} {2059} (\bibinfo {year} {2013})}\BibitemShut
  {NoStop}%
\bibitem [{\citenamefont {Anders}\ and\ \citenamefont {Esposito}(2017)}]{An17}%
  \BibitemOpen
  \bibfield  {author} {\bibinfo {author} {\bibfnamefont {J.}~\bibnamefont
  {Anders}}\ and\ \bibinfo {author} {\bibfnamefont {M.}~\bibnamefont
  {Esposito}},\ }\href {http://stacks.iop.org/1367-2630/19/i=1/a=010201}
  {\bibfield  {journal} {\bibinfo  {journal} {New Journal of Physics}\ }\textbf
  {\bibinfo {volume} {19}},\ \bibinfo {pages} {010201} (\bibinfo {year}
  {2017})}\BibitemShut {NoStop}%
\bibitem [{\citenamefont {Hofer}\ \emph {et~al.}(2017)\citenamefont {Hofer},
  \citenamefont {Brask}, \citenamefont {Perarnau-Llobet},\ and\ \citenamefont
  {Brunner}}]{Hofer17}%
  \BibitemOpen
  \bibfield  {author} {\bibinfo {author} {\bibfnamefont {P.~P.}\ \bibnamefont
  {Hofer}}, \bibinfo {author} {\bibfnamefont {J.~B.}\ \bibnamefont {Brask}},
  \bibinfo {author} {\bibfnamefont {M.}~\bibnamefont {Perarnau-Llobet}}, \ and\
  \bibinfo {author} {\bibfnamefont {N.}~\bibnamefont {Brunner}},\ }\href
  {\doibase 10.1103/PhysRevLett.119.090603} {\bibfield  {journal} {\bibinfo
  {journal} {Phys. Rev. Lett.}\ }\textbf {\bibinfo {volume} {119}},\ \bibinfo
  {pages} {090603} (\bibinfo {year} {2017})}\BibitemShut {NoStop}%
\bibitem [{\citenamefont {Gorin}\ \emph {et~al.}(2016)\citenamefont {Gorin},
  \citenamefont {Moreno},\ and\ \citenamefont {Seligman}}]{GMS16RTSA}%
  \BibitemOpen
  \bibfield  {author} {\bibinfo {author} {\bibfnamefont {T.}~\bibnamefont
  {Gorin}}, \bibinfo {author} {\bibfnamefont {H.~J.}\ \bibnamefont {Moreno}}, \
  and\ \bibinfo {author} {\bibfnamefont {T.~H.}\ \bibnamefont {Seligman}},\
  }\href {\doibase 10.1098/rsta.2015.0162} {\bibfield  {journal} {\bibinfo
  {journal} {Philos. Trans. Royal Soc. A}\ }\textbf {\bibinfo {volume}
  {374}},\ \bibinfo {pages} {20150162} (\bibinfo {year} {2016})}\BibitemShut
  {NoStop}%
\bibitem [{\citenamefont {Prado~Reynoso}\ \emph {et~al.}(2017)\citenamefont
  {Prado~Reynoso}, \citenamefont {L\'opez~V\'azquez},\ and\ \citenamefont
  {Gorin}}]{PLG17}%
  \BibitemOpen
  \bibfield  {author} {\bibinfo {author} {\bibfnamefont {M.~A.}\ \bibnamefont
  {Prado~Reynoso}}, \bibinfo {author} {\bibfnamefont {P.~C.}\ \bibnamefont
  {L\'opez~V\'azquez}}, \ and\ \bibinfo {author} {\bibfnamefont
  {T.}~\bibnamefont {Gorin}},\ }\href {\doibase 10.1103/PhysRevA.95.022118}
  {\bibfield  {journal} {\bibinfo  {journal} {Phys. Rev. A}\ }\textbf {\bibinfo
  {volume} {95}},\ \bibinfo {pages} {022118} (\bibinfo {year}
  {2017})}\BibitemShut {NoStop}%
\bibitem [{\citenamefont {Palma}\ \emph {et~al.}(1996)\citenamefont {Palma},
  \citenamefont {Suominen},\ and\ \citenamefont {Ekertand}}]{Palma96}%
  \BibitemOpen
  \bibfield  {author} {\bibinfo {author} {\bibfnamefont {G.~M.}\ \bibnamefont
  {Palma}}, \bibinfo {author} {\bibfnamefont {K.-A.}\ \bibnamefont {Suominen}},
  \ and\ \bibinfo {author} {\bibfnamefont {A.~K.}\ \bibnamefont {Ekertand}},\
  }\href {\doibase 10.1098/rspa.1996.0029} {\bibfield  {journal} {\bibinfo
  {journal} {Proc R Soc Lond A Math Phys Sci}\ }\textbf {\bibinfo {volume}
  {452}},\ \bibinfo {pages} {567} (\bibinfo {year} {1996})}\BibitemShut
  {NoStop}%
\bibitem [{\citenamefont {Reina}\ \emph {et~al.}(2002)\citenamefont {Reina},
  \citenamefont {Quiroga},\ and\ \citenamefont {Johnson}}]{Reina02}%
  \BibitemOpen
  \bibfield  {author} {\bibinfo {author} {\bibfnamefont {J.~H.}\ \bibnamefont
  {Reina}}, \bibinfo {author} {\bibfnamefont {L.}~\bibnamefont {Quiroga}}, \
  and\ \bibinfo {author} {\bibfnamefont {N.~F.}\ \bibnamefont {Johnson}},\
  }\href {\doibase 10.1103/PhysRevA.65.032326} {\bibfield  {journal} {\bibinfo
  {journal} {Phys. Rev. A}\ }\textbf {\bibinfo {volume} {65}},\ \bibinfo
  {pages} {032326} (\bibinfo {year} {2002})}\BibitemShut {NoStop}%
\bibitem [{\citenamefont {van~der Wal}\ \emph {et~al.}(2003)\citenamefont
  {van~der Wal}, \citenamefont {Wilhelm}, \citenamefont {Harmans},\ and\
  \citenamefont {Mooij}}]{vanderWal2003}%
  \BibitemOpen
  \bibfield  {author} {\bibinfo {author} {\bibfnamefont {C.}~\bibnamefont
  {van~der Wal}}, \bibinfo {author} {\bibfnamefont {F.}~\bibnamefont
  {Wilhelm}}, \bibinfo {author} {\bibfnamefont {C.}~\bibnamefont {Harmans}}, \
  and\ \bibinfo {author} {\bibfnamefont {J.}~\bibnamefont {Mooij}},\ }\href
  {\doibase 10.1140/epjb/e2003-00015-9} {\bibfield  {journal} {\bibinfo
  {journal} {Eur. Phys. J. B}\ }\textbf {\bibinfo {volume} {31}},\ \bibinfo {pages} {111}
  (\bibinfo {year} {2003})}\BibitemShut {NoStop}%
\bibitem [{\citenamefont {Brito}\ and\ \citenamefont
  {Werlang}(2015)}]{Brito15}%
  \BibitemOpen
  \bibfield  {author} {\bibinfo {author} {\bibfnamefont {F.}~\bibnamefont
  {Brito}}\ and\ \bibinfo {author} {\bibfnamefont {T.}~\bibnamefont
  {Werlang}},\ }\href {http://stacks.iop.org/1367-2630/17/i=7/a=072001}
  {\bibfield  {journal} {\bibinfo  {journal} {New Journal of Physics}\ }\textbf
  {\bibinfo {volume} {17}},\ \bibinfo {pages} {072001} (\bibinfo {year}
  {2015})}\BibitemShut {NoStop}%
\bibitem [{\citenamefont {Costa}\ \emph {et~al.}(2016)\citenamefont {Costa},
  \citenamefont {Beims},\ and\ \citenamefont {Strunz}}]{Costa16}%
  \BibitemOpen
  \bibfield  {author} {\bibinfo {author} {\bibfnamefont {A.~C.~S.}\
  \bibnamefont {Costa}}, \bibinfo {author} {\bibfnamefont {M.~W.}\ \bibnamefont
  {Beims}}, \ and\ \bibinfo {author} {\bibfnamefont {W.~T.}\ \bibnamefont
  {Strunz}},\ }\href {\doibase 10.1103/PhysRevA.93.052316} {\bibfield
  {journal} {\bibinfo  {journal} {Phys. Rev. A}\ }\textbf {\bibinfo {volume}
  {93}},\ \bibinfo {pages} {052316} (\bibinfo {year} {2016})}\BibitemShut
  {NoStop}%
\bibitem [{\citenamefont {Wallraff}\ \emph {et~al.}(2004)\citenamefont
  {Wallraff}, \citenamefont {Schuster}, \citenamefont {Blais}, \citenamefont
  {Frunzio}, \citenamefont {Huang}, \citenamefont {Majer}, \citenamefont
  {Kumar}, \citenamefont {Girvin},\ and\ \citenamefont {Schoelkopf}}]{Wa04}%
  \BibitemOpen
  \bibfield  {author} {\bibinfo {author} {\bibfnamefont {A.}~\bibnamefont
  {Wallraff}}, \bibinfo {author} {\bibfnamefont {D.~I.}\ \bibnamefont
  {Schuster}}, \bibinfo {author} {\bibfnamefont {A.}~\bibnamefont {Blais}},
  \bibinfo {author} {\bibfnamefont {L.}~\bibnamefont {Frunzio}}, \bibinfo
  {author} {\bibfnamefont {R.-S.}\ \bibnamefont {Huang}}, \bibinfo {author}
  {\bibfnamefont {J.}~\bibnamefont {Majer}}, \bibinfo {author} {\bibfnamefont
  {S.}~\bibnamefont {Kumar}}, \bibinfo {author} {\bibfnamefont {S.~M.}\
  \bibnamefont {Girvin}}, \ and\ \bibinfo {author} {\bibfnamefont {R.~J.}\
  \bibnamefont {Schoelkopf}},\ }\href {\doibase 10.1038/nature02851} {\bibfield
   {journal} {\bibinfo  {journal} {Nature}\ }\textbf {\bibinfo {volume}
  {431}},\ \bibinfo {pages} {162} (\bibinfo {year} {2004})}\BibitemShut
  {NoStop}%
\bibitem [{\citenamefont {Chiorescu}\ \emph {et~al.}(2004)\citenamefont
  {Chiorescu}, \citenamefont {Bertet}, \citenamefont {Semba}, \citenamefont
  {Nakamura}, \citenamefont {Harmans},\ and\ \citenamefont {Mooij}}]{Ch04}%
  \BibitemOpen
  \bibfield  {author} {\bibinfo {author} {\bibfnamefont {I.}~\bibnamefont
  {Chiorescu}}, \bibinfo {author} {\bibfnamefont {P.}~\bibnamefont {Bertet}},
  \bibinfo {author} {\bibfnamefont {K.}~\bibnamefont {Semba}}, \bibinfo
  {author} {\bibfnamefont {Y.}~\bibnamefont {Nakamura}}, \bibinfo {author}
  {\bibfnamefont {C.~J. P.~M.}\ \bibnamefont {Harmans}}, \ and\ \bibinfo
  {author} {\bibfnamefont {J.~E.}\ \bibnamefont {Mooij}},\ }\href {\doibase
  10.1038/nature02831} {\bibfield  {journal} {\bibinfo  {journal} {Nature}\
  }\textbf {\bibinfo {volume} {431}},\ \bibinfo {pages} {159} (\bibinfo {year}
  {2004})}\BibitemShut {NoStop}%
\bibitem [{\citenamefont {Harris}\ \emph {et~al.}(2010)\citenamefont {Harris},
  \citenamefont {Johnson}, \citenamefont {Lanting}, \citenamefont {Berkley},
  \citenamefont {Johansson}, \citenamefont {Bunyk}, \citenamefont {Tolkacheva},
  \citenamefont {Ladizinsky}, \citenamefont {Ladizinsky}, \citenamefont {Oh},
  \citenamefont {Cioata}, \citenamefont {Perminov}, \citenamefont {Spear},
  \citenamefont {Enderud}, \citenamefont {Rich}, \citenamefont {Uchaikin},
  \citenamefont {Thom}, \citenamefont {Chapple}, \citenamefont {Wang},
  \citenamefont {Wilson}, \citenamefont {Amin}, \citenamefont {Dickson},
  \citenamefont {Karimi}, \citenamefont {Macready}, \citenamefont {Truncik},\
  and\ \citenamefont {Rose}}]{Harris10}%
  \BibitemOpen
  \bibfield  {author} {\bibinfo {author} {\bibfnamefont {R.}~\bibnamefont
  {Harris}}, \bibinfo {author} {\bibfnamefont {M.~W.}\ \bibnamefont {Johnson}},
  \bibinfo {author} {\bibfnamefont {T.}~\bibnamefont {Lanting}}, \bibinfo
  {author} {\bibfnamefont {A.~J.}\ \bibnamefont {Berkley}}, \bibinfo {author}
  {\bibfnamefont {J.}~\bibnamefont {Johansson}}, \bibinfo {author}
  {\bibfnamefont {P.}~\bibnamefont {Bunyk}}, \bibinfo {author} {\bibfnamefont
  {E.}~\bibnamefont {Tolkacheva}}, \bibinfo {author} {\bibfnamefont
  {E.}~\bibnamefont {Ladizinsky}}, \bibinfo {author} {\bibfnamefont
  {N.}~\bibnamefont {Ladizinsky}}, \bibinfo {author} {\bibfnamefont
  {T.}~\bibnamefont {Oh}}, \bibinfo {author} {\bibfnamefont {F.}~\bibnamefont
  {Cioata}}, \bibinfo {author} {\bibfnamefont {I.}~\bibnamefont {Perminov}},
  \bibinfo {author} {\bibfnamefont {P.}~\bibnamefont {Spear}}, \bibinfo
  {author} {\bibfnamefont {C.}~\bibnamefont {Enderud}}, \bibinfo {author}
  {\bibfnamefont {C.}~\bibnamefont {Rich}}, \bibinfo {author} {\bibfnamefont
  {S.}~\bibnamefont {Uchaikin}}, \bibinfo {author} {\bibfnamefont {M.~C.}\
  \bibnamefont {Thom}}, \bibinfo {author} {\bibfnamefont {E.~M.}\ \bibnamefont
  {Chapple}}, \bibinfo {author} {\bibfnamefont {J.}~\bibnamefont {Wang}},
  \bibinfo {author} {\bibfnamefont {B.}~\bibnamefont {Wilson}}, \bibinfo
  {author} {\bibfnamefont {M.~H.~S.}\ \bibnamefont {Amin}}, \bibinfo {author}
  {\bibfnamefont {N.}~\bibnamefont {Dickson}}, \bibinfo {author} {\bibfnamefont
  {K.}~\bibnamefont {Karimi}}, \bibinfo {author} {\bibfnamefont
  {B.}~\bibnamefont {Macready}}, \bibinfo {author} {\bibfnamefont {C.~J.~S.}\
  \bibnamefont {Truncik}}, \ and\ \bibinfo {author} {\bibfnamefont
  {G.}~\bibnamefont {Rose}},\ }\href {\doibase 10.1103/PhysRevB.82.024511}
  {\bibfield  {journal} {\bibinfo  {journal} {Phys. Rev. B}\ }\textbf {\bibinfo
  {volume} {82}},\ \bibinfo {pages} {024511} (\bibinfo {year}
  {2010})}\BibitemShut {NoStop}%
\bibitem [{\citenamefont {Friedenauer}\ \emph {et~al.}(2008)\citenamefont
  {Friedenauer}, \citenamefont {Schmitz}, \citenamefont {Glueckert},\ and\
  \citenamefont {Porras}}]{Fri08}%
  \BibitemOpen
  \bibfield  {author} {\bibinfo {author} {\bibfnamefont {A.}~\bibnamefont
  {Friedenauer}}, \bibinfo {author} {\bibfnamefont {H.}~\bibnamefont
  {Schmitz}}, \bibinfo {author} {\bibfnamefont {J.~T.}\ \bibnamefont
  {Glueckert}}, \ and\ \bibinfo {author} {\bibfnamefont {T.}~\bibnamefont
  {Porras}, \bibfnamefont {D.and~Schaetz}},\ }\href {\doibase
  10.1038/nphys1032} {\bibfield  {journal} {\bibinfo  {journal} {Nature
  Physics}\ }\textbf {\bibinfo {volume} {4}},\ \bibinfo {pages} {757} (\bibinfo
  {year} {2008})}\BibitemShut {NoStop}%
\bibitem [{\citenamefont {Porras}\ \emph {et~al.}(2008)\citenamefont {Porras},
  \citenamefont {Marquardt}, \citenamefont {von Delft},\ and\ \citenamefont
  {Cirac}}]{Porras08}%
  \BibitemOpen
  \bibfield  {author} {\bibinfo {author} {\bibfnamefont {D.}~\bibnamefont
  {Porras}}, \bibinfo {author} {\bibfnamefont {F.}~\bibnamefont {Marquardt}},
  \bibinfo {author} {\bibfnamefont {J.}~\bibnamefont {von Delft}}, \ and\
  \bibinfo {author} {\bibfnamefont {J.~I.}\ \bibnamefont {Cirac}},\ }\href
  {\doibase 10.1103/PhysRevA.78.010101} {\bibfield  {journal} {\bibinfo
  {journal} {Phys. Rev. A}\ }\textbf {\bibinfo {volume} {78}},\ \bibinfo
  {pages} {010101} (\bibinfo {year} {2008})}\BibitemShut {NoStop}%
\bibitem [{\citenamefont {Kim}\ \emph {et~al.}(2011)\citenamefont {Kim},
  \citenamefont {Korenblit}, \citenamefont {Islam}, \citenamefont {Edwards},
  \citenamefont {Chang}, \citenamefont {Noh}, \citenamefont {Carmichael},
  \citenamefont {Lin}, \citenamefont {Duan}, \citenamefont {Wang},
  \citenamefont {Freericks},\ and\ \citenamefont {Monroe}}]{Kim11}%
  \BibitemOpen
  \bibfield  {author} {\bibinfo {author} {\bibfnamefont {K.}~\bibnamefont
  {Kim}}, \bibinfo {author} {\bibfnamefont {S.}~\bibnamefont {Korenblit}},
  \bibinfo {author} {\bibfnamefont {R.}~\bibnamefont {Islam}}, \bibinfo
  {author} {\bibfnamefont {E.~E.}\ \bibnamefont {Edwards}}, \bibinfo {author}
  {\bibfnamefont {M.-S.}\ \bibnamefont {Chang}}, \bibinfo {author}
  {\bibfnamefont {C.}~\bibnamefont {Noh}}, \bibinfo {author} {\bibfnamefont
  {H.}~\bibnamefont {Carmichael}}, \bibinfo {author} {\bibfnamefont {G.-D.}\
  \bibnamefont {Lin}}, \bibinfo {author} {\bibfnamefont {L.-M.}\ \bibnamefont
  {Duan}}, \bibinfo {author} {\bibfnamefont {C.~C.~J.}\ \bibnamefont {Wang}},
  \bibinfo {author} {\bibfnamefont {J.~K.}\ \bibnamefont {Freericks}}, \ and\
  \bibinfo {author} {\bibfnamefont {C.}~\bibnamefont {Monroe}},\ }\href
  {http://stacks.iop.org/1367-2630/13/i=10/a=105003} {\bibfield  {journal}
  {\bibinfo  {journal} {New Journal of Physics}\ }\textbf {\bibinfo {volume}
  {13}},\ \bibinfo {pages} {105003} (\bibinfo {year} {2011})}\BibitemShut
  {NoStop}%
\bibitem [{\citenamefont {Schindler}\ \emph {et~al.}(2013)\citenamefont
  {Schindler}, \citenamefont {Müller}, \citenamefont {Nigg}, \citenamefont
  {Martinez}, \citenamefont {Hennrich}, \citenamefont {Monz}, \citenamefont
  {Diehl}, \citenamefont {Zoller},\ and\ \citenamefont {Blatt}}]{Sch13}%
  \BibitemOpen
  \bibfield  {author} {\bibinfo {author} {\bibfnamefont {P.}~\bibnamefont
  {Schindler}}, \bibinfo {author} {\bibfnamefont {M.}~\bibnamefont {Müller}},
  \bibinfo {author} {\bibfnamefont {J.~T.}\ \bibnamefont {Nigg}, \bibfnamefont
  {D.and~Barreiro}}, \bibinfo {author} {\bibfnamefont {E.~A.}\ \bibnamefont
  {Martinez}}, \bibinfo {author} {\bibfnamefont {M.}~\bibnamefont {Hennrich}},
  \bibinfo {author} {\bibfnamefont {T.}~\bibnamefont {Monz}}, \bibinfo {author}
  {\bibfnamefont {S.}~\bibnamefont {Diehl}}, \bibinfo {author} {\bibfnamefont
  {P.}~\bibnamefont {Zoller}}, \ and\ \bibinfo {author} {\bibfnamefont
  {R.}~\bibnamefont {Blatt}},\ }\href {\doibase 10.1038/nphys2630} {\bibfield
  {journal} {\bibinfo  {journal} {Nature Physics}\ }\textbf {\bibinfo {volume}
  {9}},\ \bibinfo {pages} {361} (\bibinfo {year} {2013})}\BibitemShut {NoStop}%
\bibitem [{\citenamefont {Simon}\ \emph {et~al.}(2011)\citenamefont {Simon},
  \citenamefont {Bakr}, \citenamefont {Ma}, \citenamefont {Tai},\ and\
  \citenamefont {Preiss}}]{Sim11}%
  \BibitemOpen
  \bibfield  {author} {\bibinfo {author} {\bibfnamefont {J.}~\bibnamefont
  {Simon}}, \bibinfo {author} {\bibfnamefont {W.~S.}\ \bibnamefont {Bakr}},
  \bibinfo {author} {\bibfnamefont {R.}~\bibnamefont {Ma}}, \bibinfo {author}
  {\bibfnamefont {M.~E.}\ \bibnamefont {Tai}}, \ and\ \bibinfo {author}
  {\bibfnamefont {M.}~\bibnamefont {Preiss}, \bibfnamefont {Philipp
  M.and~Greiner}},\ }\href {\doibase 10.1038/nature09994} {\bibfield  {journal}
  {\bibinfo  {journal} {Nature}\ }\textbf {\bibinfo {volume} {472}},\ \bibinfo
  {pages} {307} (\bibinfo {year} {2011})}\BibitemShut {NoStop}%
\bibitem [{\citenamefont {Recati}\ \emph {et~al.}(2005)\citenamefont {Recati},
  \citenamefont {Fedichev}, \citenamefont {Zwerger}, \citenamefont {von
  Delft},\ and\ \citenamefont {Zoller}}]{Recati05}%
  \BibitemOpen
  \bibfield  {author} {\bibinfo {author} {\bibfnamefont {A.}~\bibnamefont
  {Recati}}, \bibinfo {author} {\bibfnamefont {P.~O.}\ \bibnamefont
  {Fedichev}}, \bibinfo {author} {\bibfnamefont {W.}~\bibnamefont {Zwerger}},
  \bibinfo {author} {\bibfnamefont {J.}~\bibnamefont {von Delft}}, \ and\
  \bibinfo {author} {\bibfnamefont {P.}~\bibnamefont {Zoller}},\ }\href
  {\doibase 10.1103/PhysRevLett.94.040404} {\bibfield  {journal} {\bibinfo
  {journal} {Phys. Rev. Lett.}\ }\textbf {\bibinfo {volume} {94}},\ \bibinfo
  {pages} {040404} (\bibinfo {year} {2005})}\BibitemShut {NoStop}%
\bibitem [{\citenamefont {Orth}\ \emph {et~al.}(2008)\citenamefont {Orth},
  \citenamefont {Stanic},\ and\ \citenamefont {Le~Hur}}]{Orth08}%
  \BibitemOpen
  \bibfield  {author} {\bibinfo {author} {\bibfnamefont {P.~P.}\ \bibnamefont
  {Orth}}, \bibinfo {author} {\bibfnamefont {I.}~\bibnamefont {Stanic}}, \ and\
  \bibinfo {author} {\bibfnamefont {K.}~\bibnamefont {Le~Hur}},\ }\href
  {\doibase 10.1103/PhysRevA.77.051601} {\bibfield  {journal} {\bibinfo
  {journal} {Phys. Rev. A}\ }\textbf {\bibinfo {volume} {77}},\ \bibinfo
  {pages} {051601} (\bibinfo {year} {2008})}\BibitemShut {NoStop}%
\bibitem [{\citenamefont {Haikka}\ \emph {et~al.}(2011)\citenamefont {Haikka},
  \citenamefont {McEndoo}, \citenamefont {De~Chiara}, \citenamefont {Palma},\
  and\ \citenamefont {Maniscalco}}]{Haikka11}%
  \BibitemOpen
  \bibfield  {author} {\bibinfo {author} {\bibfnamefont {P.}~\bibnamefont
  {Haikka}}, \bibinfo {author} {\bibfnamefont {S.}~\bibnamefont {McEndoo}},
  \bibinfo {author} {\bibfnamefont {G.}~\bibnamefont {De~Chiara}}, \bibinfo
  {author} {\bibfnamefont {G.~M.}\ \bibnamefont {Palma}}, \ and\ \bibinfo
  {author} {\bibfnamefont {S.}~\bibnamefont {Maniscalco}},\ }\href {\doibase
  10.1103/PhysRevA.84.031602} {\bibfield  {journal} {\bibinfo  {journal} {Phys.
  Rev. A}\ }\textbf {\bibinfo {volume} {84}},\ \bibinfo {pages} {031602}
  (\bibinfo {year} {2011})}\BibitemShut {NoStop}%
\bibitem [{\citenamefont {Makhlin}\ \emph {et~al.}(2001)\citenamefont
  {Makhlin}, \citenamefont {Sch\"on},\ and\ \citenamefont {Shnirman}}]{Ma01}%
  \BibitemOpen
  \bibfield  {author} {\bibinfo {author} {\bibfnamefont {Y.}~\bibnamefont
  {Makhlin}}, \bibinfo {author} {\bibfnamefont {G.}~\bibnamefont {Sch\"on}}, \
  and\ \bibinfo {author} {\bibfnamefont {A.}~\bibnamefont {Shnirman}},\ }\href
  {\doibase 10.1103/RevModPhys.73.357} {\bibfield  {journal} {\bibinfo
  {journal} {Rev. Mod. Phys.}\ }\textbf {\bibinfo {volume} {73}},\ \bibinfo
  {pages} {357} (\bibinfo {year} {2001})}\BibitemShut {NoStop}%
\bibitem [{\citenamefont {Sornborger}\ \emph {et~al.}(2004)\citenamefont
  {Sornborger}, \citenamefont {Cleland},\ and\ \citenamefont {Geller}}]{So04}%
  \BibitemOpen
  \bibfield  {author} {\bibinfo {author} {\bibfnamefont {A.~T.}\ \bibnamefont
  {Sornborger}}, \bibinfo {author} {\bibfnamefont {A.~N.}\ \bibnamefont
  {Cleland}}, \ and\ \bibinfo {author} {\bibfnamefont {M.~R.}\ \bibnamefont
  {Geller}},\ }\href {\doibase 10.1103/PhysRevA.70.052315} {\bibfield
  {journal} {\bibinfo  {journal} {Phys. Rev. A}\ }\textbf {\bibinfo {volume}
  {70}},\ \bibinfo {pages} {052315} (\bibinfo {year} {2004})}\BibitemShut
  {NoStop}%
\bibitem [{\citenamefont {Betzholz}\ \emph {et~al.}(2014)\citenamefont
  {Betzholz}, \citenamefont {Torres},\ and\ \citenamefont
  {Bienert}}]{BeToBi14}%
  \BibitemOpen
  \bibfield  {author} {\bibinfo {author} {\bibfnamefont {R.}~\bibnamefont
  {Betzholz}}, \bibinfo {author} {\bibfnamefont {J.~M.}\ \bibnamefont
  {Torres}}, \ and\ \bibinfo {author} {\bibfnamefont {M.}~\bibnamefont
  {Bienert}},\ }\href {\doibase 10.1103/PhysRevA.90.063818} {\bibfield
  {journal} {\bibinfo  {journal} {Phys. Rev. A}\ }\textbf {\bibinfo {volume}
  {90}},\ \bibinfo {pages} {063818} (\bibinfo {year} {2014})}\BibitemShut
  {NoStop}%
\bibitem [{\citenamefont {Scutaru}(1998)}]{Scu98}%
  \BibitemOpen
  \bibfield  {author} {\bibinfo {author} {\bibfnamefont {H.}~\bibnamefont
  {Scutaru}},\ }\href {http://stacks.iop.org/0305-4470/31/i=15/a=025}
  {\bibfield  {journal} {\bibinfo  {journal} {J Phys A Math Theor}\ }
  \textbf {\bibinfo {volume} {31}},\ \bibinfo
  {pages} {3659} (\bibinfo {year} {1998})}\BibitemShut {NoStop}%
\bibitem [{\citenamefont {Isar}(2009)}]{Isar09}%
  \BibitemOpen
  \bibfield  {author} {\bibinfo {author} {\bibfnamefont {A.}~\bibnamefont
  {Isar}},\ }\href {\doibase 10.1134/S1547477109070164} {\bibfield  {journal}
  {\bibinfo  {journal} {Physics of Particles and Nuclei Letters}\ }\textbf
  {\bibinfo {volume} {6}},\ \bibinfo {pages} {567} (\bibinfo {year}
  {2009})}\BibitemShut {NoStop}%
\bibitem [{\citenamefont {Uhlmann}(1976)}]{Uhl76}%
  \BibitemOpen
  \bibfield  {author} {\bibinfo {author} {\bibfnamefont {A.}~\bibnamefont
  {Uhlmann}},\ }\href@noop {} {\bibfield  {journal} {\bibinfo  {journal}
  {Reports on Mathematical Physics}\ }\textbf {\bibinfo {volume} {9}},\
  \bibinfo {pages} {273} (\bibinfo {year} {1976})}\BibitemShut {NoStop}%
\bibitem [{\citenamefont {Jozsa}(1994)}]{Joz94}%
  \BibitemOpen
  \bibfield  {author} {\bibinfo {author} {\bibfnamefont {R.}~\bibnamefont
  {Jozsa}},\ }\href {\doibase 10.1080/09500349414552171} {\bibfield  {journal}
  {\bibinfo  {journal} {Journal of Modern Optics}\ }\textbf {\bibinfo {volume}
  {41}},\ \bibinfo {pages} {2315} (\bibinfo {year} {1994})}\BibitemShut
  {NoStop}%
\bibitem [{\citenamefont {Clerk}\ and\ \citenamefont {Utami}(2007)}]{Clerk07}%
  \BibitemOpen
  \bibfield  {author} {\bibinfo {author} {\bibfnamefont {A.~A.}\ \bibnamefont
  {Clerk}}\ and\ \bibinfo {author} {\bibfnamefont {D.~W.}\ \bibnamefont
  {Utami}},\ }\href {\doibase 10.1103/PhysRevA.75.042302} {\bibfield  {journal}
  {\bibinfo  {journal} {Phys. Rev. A}\ }\textbf {\bibinfo {volume} {75}},\
  \bibinfo {pages} {042302} (\bibinfo {year} {2007})}\BibitemShut {NoStop}%
\bibitem [{\citenamefont {Zhao}\ and\ \citenamefont {Yin}(2014)}]{Zhao14}%
  \BibitemOpen
  \bibfield  {author} {\bibinfo {author} {\bibfnamefont {N.}~\bibnamefont
  {Zhao}}\ and\ \bibinfo {author} {\bibfnamefont {Z.-q.}\ \bibnamefont {Yin}},\
  }\href {\doibase 10.1103/PhysRevA.90.042118} {\bibfield  {journal} {\bibinfo
  {journal} {Phys. Rev. A}\ }\textbf {\bibinfo {volume} {90}},\ \bibinfo
  {pages} {042118} (\bibinfo {year} {2014})}\BibitemShut {NoStop}%
\bibitem [{\citenamefont {Jagadish}\ and\ \citenamefont {Shaji}(2015)}]{Vin15}%
  \BibitemOpen
  \bibfield  {author} {\bibinfo {author} {\bibfnamefont {V.}~\bibnamefont
  {Jagadish}}\ and\ \bibinfo {author} {\bibfnamefont {A.}~\bibnamefont
  {Shaji}},\ }\href {\doibase https://doi.org/10.1016/j.aop.2015.08.006}
  {\bibfield  {journal} {\bibinfo  {journal} {Annals of Physics}\ }\textbf
  {\bibinfo {volume} {362}},\ \bibinfo {pages} {287 } (\bibinfo {year}
  {2015})}\BibitemShut {NoStop}%
\bibitem [{\citenamefont {Caldeira}\ and\ \citenamefont
  {Leggett}(1983)}]{CalLeg83}%
  \BibitemOpen
  \bibfield  {author} {\bibinfo {author} {\bibfnamefont {A.~O.}\ \bibnamefont
  {Caldeira}}\ and\ \bibinfo {author} {\bibfnamefont {A.~J.}\ \bibnamefont
  {Leggett}},\ }\href@noop {} {\bibfield  {journal} {\bibinfo  {journal}
  {Physica}\ }\textbf {\bibinfo {volume} {121A}},\ \bibinfo {pages} {587}
  (\bibinfo {year} {1983})}\BibitemShut {NoStop}%
\bibitem [{\citenamefont {Gorin}\ \emph {et~al.}(2004)\citenamefont {Gorin},
  \citenamefont {Prosen}, \citenamefont {Seligman},\ and\ \citenamefont
  {Strunz}}]{GPSS04}%
  \BibitemOpen
  \bibfield  {author} {\bibinfo {author} {\bibfnamefont {T.}~\bibnamefont
  {Gorin}}, \bibinfo {author} {\bibfnamefont {T.}~\bibnamefont {Prosen}},
  \bibinfo {author} {\bibfnamefont {T.~H.}\ \bibnamefont {Seligman}}, \ and\
  \bibinfo {author} {\bibfnamefont {W.~T.}\ \bibnamefont {Strunz}},\
  }\href@noop {} {\bibfield  {journal} {\bibinfo  {journal} {Phys. Rev. A}\
  }\textbf {\bibinfo {volume} {70}},\ \bibinfo {pages} {042105:1} (\bibinfo
  {year} {2004})}\BibitemShut {NoStop}%
\bibitem [{\citenamefont {Sudarshan}\ \emph {et~al.}(1961)\citenamefont
  {Sudarshan}, \citenamefont {Mathews},\ and\ \citenamefont {Rau}}]{Sud61}%
  \BibitemOpen
  \bibfield  {author} {\bibinfo {author} {\bibfnamefont {E.~C.~G.}\
  \bibnamefont {Sudarshan}}, \bibinfo {author} {\bibfnamefont {P.~M.}\
  \bibnamefont {Mathews}}, \ and\ \bibinfo {author} {\bibfnamefont
  {J.}~\bibnamefont {Rau}},\ }\href {\doibase 10.1103/PhysRev.121.920}
  {\bibfield  {journal} {\bibinfo  {journal} {Phys. Rev.}\ }\textbf {\bibinfo
  {volume} {121}},\ \bibinfo {pages} {920} (\bibinfo {year}
  {1961})}\BibitemShut {NoStop}%
\bibitem [{\citenamefont {Gorini}\ \emph {et~al.}(1976)\citenamefont {Gorini},
  \citenamefont {Kossakowski},\ and\ \citenamefont {Sudarshan}}]{GoKoSu76}%
  \BibitemOpen
  \bibfield  {author} {\bibinfo {author} {\bibfnamefont {V.}~\bibnamefont
  {Gorini}}, \bibinfo {author} {\bibfnamefont {A.}~\bibnamefont {Kossakowski}},
  \ and\ \bibinfo {author} {\bibfnamefont {E.~C.~G.}\ \bibnamefont
  {Sudarshan}},\ }\href {\doibase http://dx.doi.org/10.1063/1.522979}
  {\bibfield  {journal} {\bibinfo  {journal} {Journal of Mathematical Physics}\
  }\textbf {\bibinfo {volume} {17}},\ \bibinfo {pages} {821} (\bibinfo {year}
  {1976})}\BibitemShut {NoStop}%
\bibitem [{\citenamefont {Lindblad}(1976)}]{Lin76}%
  \BibitemOpen
  \bibfield  {author} {\bibinfo {author} {\bibfnamefont {G.}~\bibnamefont
  {Lindblad}},\ }\href@noop {} {\bibfield  {journal} {\bibinfo  {journal}
  {Commun. Math. Phys.}\ }\textbf {\bibinfo {volume} {48}},\ \bibinfo {pages}
  {119} (\bibinfo {year} {1976})}\BibitemShut {NoStop}%
\bibitem [{\citenamefont {Moreno}\ \emph {et~al.}(2015)\citenamefont {Moreno},
  \citenamefont {Gorin},\ and\ \citenamefont {Seligman}}]{MGS15}%
  \BibitemOpen
  \bibfield  {author} {\bibinfo {author} {\bibfnamefont {H.~J.}\ \bibnamefont
  {Moreno}}, \bibinfo {author} {\bibfnamefont {T.}~\bibnamefont {Gorin}}, \
  and\ \bibinfo {author} {\bibfnamefont {T.~H.}\ \bibnamefont {Seligman}},\
  }\href {\doibase 10.1103/PhysRevA.92.030104} {\bibfield  {journal} {\bibinfo
  {journal} {Phys. Rev. A}\ }\textbf {\bibinfo {volume} {92}},\ \bibinfo
  {pages} {030104} (\bibinfo {year} {2015})}\BibitemShut {NoStop}%
\bibitem [{\citenamefont {de~Almeida}(1998)}]{Ozo98}%
  \BibitemOpen
  \bibfield  {author} {\bibinfo {author} {\bibfnamefont {A.~M.}\ \bibnamefont
  {de~Almeida}},\ }\href {\doibase
  http://dx.doi.org/10.1016/S0370-1573(97)00070-7} {\bibfield  {journal}
  {\bibinfo  {journal} {Physics Reports}\ }\textbf {\bibinfo {volume} {295}},\
  \bibinfo {pages} {265 } (\bibinfo {year} {1998})}\BibitemShut {NoStop}%
\bibitem [{\citenamefont {de~Almeida}(2003)}]{Ozo02}%
  \BibitemOpen
  \bibfield  {author} {\bibinfo {author} {\bibfnamefont {A.~M.~O.}\
  \bibnamefont {de~Almeida}},\ }\href
  {http://stacks.iop.org/0305-4470/36/i=1/a=305} {\bibfield  {journal}
  {\bibinfo  {journal} {J Phys A Math Gen}\
  }\textbf {\bibinfo {volume} {36}},\ \bibinfo {pages} {67} (\bibinfo {year}
  {2003})}\BibitemShut {NoStop}%
\bibitem [{\citenamefont {Brodier}\ and\ \citenamefont {Almeida}(2004)}]{Oz04}%
  \BibitemOpen
  \bibfield  {author} {\bibinfo {author} {\bibfnamefont {O.}~\bibnamefont
  {Brodier}}\ and\ \bibinfo {author} {\bibfnamefont {A.~M. O.~d.}\ \bibnamefont
  {Almeida}},\ }\href {\doibase 10.1103/PhysRevE.69.016204} {\bibfield
  {journal} {\bibinfo  {journal} {Phys. Rev. E}\ }\textbf {\bibinfo {volume}
  {69}},\ \bibinfo {pages} {016204} (\bibinfo {year} {2004})}\BibitemShut
  {NoStop}%
\bibitem [{\citenamefont {Breuer}\ and\ \citenamefont
  {Petruccione}(2002)}]{BrePet02}%
  \BibitemOpen
  \bibfield  {author} {\bibinfo {author} {\bibfnamefont {H.~P.}\ \bibnamefont
  {Breuer}}\ and\ \bibinfo {author} {\bibfnamefont {F.}~\bibnamefont
  {Petruccione}},\ }\href {\doibase 10.1093/acprof:oso/9780199213900.001.0001}
  {\emph {\bibinfo {title} {The Theory of Open Quantum Systems}}}\ (\bibinfo
  {publisher} {Oxford University Press, USA},\ \bibinfo {year}
  {2002})\BibitemShut {NoStop}%
\bibitem [{\citenamefont {Weyl}(1927)}]{weyl27}%
  \BibitemOpen
  \bibfield  {author} {\bibinfo {author} {\bibfnamefont {H.}~\bibnamefont
  {Weyl}},\ }\href {\doibase 10.1007/BF02055756} {\bibfield  {journal}
  {\bibinfo  {journal} {Zeitschrift f{\"u}r Physik}\ }\textbf {\bibinfo
  {volume} {46}},\ \bibinfo {pages} {1} (\bibinfo {year} {1927})}\BibitemShut
  {NoStop}%
\bibitem [{\citenamefont {Klimov}\ and\ \citenamefont {Chumakov}(2009)}]{Kl09}%
  \BibitemOpen
  \bibfield  {author} {\bibinfo {author} {\bibfnamefont {A.~B.}\ \bibnamefont
  {Klimov}}\ and\ \bibinfo {author} {\bibfnamefont {S.~M.}\ \bibnamefont
  {Chumakov}},\ }\href {\doibase 10.1002/9783527624003} {\emph {\bibinfo
  {title} {A Group-Theoretical Approach to Quantum Optics}}}\ (\bibinfo
  {publisher} {Wiley-VCH Verlag GmbH and Co. KGaA},\ \bibinfo {year}
  {2009})\BibitemShut {NoStop}%
\bibitem [{\citenamefont {Rigas}\ \emph {et~al.}(2011)\citenamefont {Rigas},
  \citenamefont {Sánchez-Soto}, \citenamefont {Klimov}, \citenamefont
  {Řeháček},\ and\ \citenamefont {Hradil}}]{Rig11}%
  \BibitemOpen
  \bibfield  {author} {\bibinfo {author} {\bibfnamefont {I.}~\bibnamefont
  {Rigas}}, \bibinfo {author} {\bibfnamefont {L.}~\bibnamefont
  {Sánchez-Soto}}, \bibinfo {author} {\bibfnamefont {A.}~\bibnamefont
  {Klimov}}, \bibinfo {author} {\bibfnamefont {J.}~\bibnamefont {Řeháček}},
  \ and\ \bibinfo {author} {\bibfnamefont {Z.}~\bibnamefont {Hradil}},\ }\href
  {\doibase https://doi.org/10.1016/j.aop.2010.11.016} {\bibfield  {journal}
  {\bibinfo  {journal} {Annals of Physics}\ }\textbf {\bibinfo {volume}
  {326}},\ \bibinfo {pages} {426 } (\bibinfo {year} {2011})}\BibitemShut
  {NoStop}%
\bibitem [{\citenamefont {Case}(2008)}]{Case08}%
  \BibitemOpen
  \bibfield  {author} {\bibinfo {author} {\bibfnamefont {W.~B.}\ \bibnamefont
  {Case}},\ }\href {\doibase 10.1119/1.2957889} {\bibfield  {journal} {\bibinfo
   {journal} {American Journal of Physics}\ }\textbf {\bibinfo {volume} {76}},\
  \bibinfo {pages} {937} (\bibinfo {year} {2008})}\BibitemShut {NoStop}%
\bibitem [{\citenamefont {Puri}\ and\ \citenamefont
  {Agarwal}(1987)}]{PurAga87}%
  \BibitemOpen
  \bibfield  {author} {\bibinfo {author} {\bibfnamefont {R.~R.}\ \bibnamefont
  {Puri}}\ and\ \bibinfo {author} {\bibfnamefont {G.~S.}\ \bibnamefont
  {Agarwal}},\ }\href {\doibase 10.1103/PhysRevA.35.3433} {\bibfield  {journal}
  {\bibinfo  {journal} {Phys. Rev. A}\ }\textbf {\bibinfo {volume} {35}},\
  \bibinfo {pages} {3433} (\bibinfo {year} {1987})}\BibitemShut {NoStop}%
\bibitem [{\citenamefont {Haug}\ \emph {et~al.}(2005)\citenamefont {Haug},
  \citenamefont {Bienert}, \citenamefont {Schleich}, \citenamefont {Seligman},\
  and\ \citenamefont {Raizen}}]{Haug05}%
  \BibitemOpen
  \bibfield  {author} {\bibinfo {author} {\bibfnamefont {F.}~\bibnamefont
  {Haug}}, \bibinfo {author} {\bibfnamefont {M.}~\bibnamefont {Bienert}},
  \bibinfo {author} {\bibfnamefont {W.~P.}\ \bibnamefont {Schleich}}, \bibinfo
  {author} {\bibfnamefont {T.~H.}\ \bibnamefont {Seligman}}, \ and\ \bibinfo
  {author} {\bibfnamefont {M.~G.}\ \bibnamefont {Raizen}},\ }\href@noop {}
  {\bibfield  {journal} {\bibinfo  {journal} {Phys. Rev. A}\ }\textbf {\bibinfo
  {volume} {71}},\ \bibinfo {pages} {043803:1} (\bibinfo {year}
  {2005})}\BibitemShut {NoStop}%
\end{thebibliography}%

\end{document}